\documentclass[pra,preprint,superscriptaddress,floatfix,showpacs]{revtex4-1}

\usepackage{amsmath,amsfonts,amssymb}
\usepackage{graphicx,color}

\def\beq{\begin{equation}}
\def\eeq{\end{equation}}
\def\beqn{\begin{eqnarray}}
\def\eeqn{\end{eqnarray}}

\def\r {{\bf r}}

\def\r {{\bf r}}

\begin{document}

\title{Condensates in annuli: Dimensionality of the variance}
\author{Ofir E. Alon}
\email{ofir@research.haifa.ac.il}
\affiliation{Department of Mathematics, University of Haifa, Haifa 3498838, Israel}
\affiliation{Haifa Research Center for Theoretical Physics and Astrophysics, University of Haifa,
Haifa 3498838, Israel}

\begin{abstract}
Static and dynamic properties of Bose-Einstein 
condensates in annular traps
are investigated
by solving the many-boson Schr\"odinger equation numerically accurately
using the multiconfigurational time-dependent Hartree for bosons
method.
We concentrate on weakly-interacting bosons exhibiting low depletion.
Analysis of the mean-field position variance, which accounts for the shape of the
density only,
and the many-body position variance, which incorporates a tiny amount of excitations through the reduced two-particle density matrix,
shows that the former behaves essentially as a quasi-one-dimensional quantity
whereas the latter as a two-dimensional quantity.
This brings another dimension to
the physics of bosons in
ring-shaped traps.
\end{abstract}


\maketitle 

\section{Introduction}\label{INTRODUCTION}

The realization of Bose-Einstein condensates (BECs) with 
ultra-cold atoms in magnetic and optical traps has opened a venue for comparing
theory and experiment in a well-engineered manner \cite{rev1,rev2,rev3,rev4,rev5}.
Within theory itself, the inter-connection between mean-field and many-body descriptions of a BEC
has drawn much attention.
Whereas mean-field, or, as it is also known, Gross-Pitaevskii theory has widely been employed
in earlier investigations,
there is nowadays a growing consensus of the need to often go beyond mean field.
Here, exact and appealing relations between many-body and mean-field
theories of a BEC can be made in the so-called infinite-particle limit
(the infinite-particle limit is defined such that the interaction parameter, i.e., the product of the interaction 
strength times the number of particles $N$, is kept fixed for increasing number of particles)
\cite{INF1,INF2,INF3,INF4,INF5,INF6}.
Explicitly, the energy per particle, $\frac{E}{N}$, and density per particle, $\frac{\rho(\r)}{N}$, of the BEC 
computed at the many-body and mean-field levels of theory for $N \to \infty$ are equal. 
Moreover, the BEC is $100\%$ condensed,
a result obtained for {\it any} finite-order reduced density matrix.

Even then, there are correlations embedded within a BEC.
Indeed, the variance of a one-body operator, like the position operator, 
$\frac{1}{N}\Delta^2_{\hat X} = \frac{1}{N}\left(\langle\hat X^2\rangle - \langle\hat X\rangle^2\right)$ with 
$\hat X=\sum_{j=1}^N \hat x_j$, where $\hat x_j$ is the position of the $j$th particle,
can extract such correlations \cite{var1,var2}.
The reason lies in the excitation of as little as a fraction of a particle outside the condensed mode,
which then interacts with the macroscopic number of particles in the condensed mode.
Formally, in the evaluation of the variance of one-particle operators two-particle operators contribute,
$\hat X^2=\sum_{j=1}^N \hat x^2_j + \sum_{k>j=1}^N 2\hat x_j \hat x_k$.
This is an intriguing result,
especially since both the reduced one- and two-particle density matrices
are $100\%$ condensed at the infinite-particle limit.

In practice, one finds a difference when the variance is computed at the many-body and mean-field levels,
and can investigate how this difference is associated with correlations in BECs.
A broader facet to be mentioned is the relation between the many-body and mean-field wavefunctions themselves.
In \cite{overlap}, it has been shown that the overlap 
of the many-body and mean-field wavefunctions can become (much) smaller than $1$.
In turn, even at the infinite-particle limit the many-body wavefunction is extremely complex and very different from the mean-field 
one and this is caused by the very few bosons outside the condensed mode \cite{INF6}.
When the variance is computed at the mean-field level, i.e., from the mean-field wavefunction,
it directly relates to the shape or size of the density.
When the variance is computed from the many-body wavefunction it incorporates correlations,
and can be associated with an effective `many-body' shape or size of the BEC.
This opens a door for studying the relation between the shape of a BEC
(given by its density) and the correlations
within a BEC embedded in its many-body variance.

The difference between the many-body variance and the shape of a BEC 
depends on the strength of the interaction between bosons, 
the geometry of the trap holding the bosons, and the observable under investigation,
e.g., the position, momentum, or angular momentum \cite{var1,var2,var3}.
In dynamical scenarios, it also depends on time.
For instance, a repulsive interaction between bosons leads to a broader density,
whereas the many-body variance is actually smaller. 
This result holds for the ground state as well as for the time-dependent many-body variance 
in an interaction-quench scenario.
Since the density and many-body variance of a BEC can behave in an opposite manner,
the question what does happen in traps of different topologies becomes interesting.
Here, it has recently been shown that the density and many-body variance in an anisotropic 
trap can exhibit opposite anisotropies, i.e.,
if the density along the $y$ direction is wider than along the $x$ direction,
then the many-body variance along the $y$ direction is smaller than along the $x$ direction \cite{var4}.
On the applications side,
the many-body variance of BECs can be used as a sensitive diagnostic tool,
for excitations of BECs \cite{var_ap1} (also see \cite{rapha_excite}),
for analyzing the impact of the range of interaction \cite{var_ap2},
and for assessing convergence of numerical simulations \cite{brand_1,attractive_2}.

So far, the connection between the many-body variance and shape of a BEC
was investigated in simply-connected traps.
In this work we consider a non-simply-connected trap, a ring in two spatial dimensions, or, simply, an annulus, 
and study the properties of interacting bosons loaded in it.
The focus of investigations is the many-body variance,
but we also discuss for our needs the energy and depletion in the system.
Both statics and dynamics in the annulus are investigated.
Bosons in rings and annuli have drawn much attention
\cite{rn1,rn2,rn3,rn4,rn5,rn6,rn7,rn8,rn9,rn10,rn11,rn12,rn13,rn14,rn15,rn16,rn17,rn18,rn19,rn20,rn21,rn22}
in one-, two-, and three-dimensional setups.
Another non-simply-connected geometry of interest is 
hollow BECs in three spatial dimensions 
which were studied too \cite{hol1,hol2,hol3}.

The structure of the paper is as follows.
In Sec.~\ref{THEORY} we present the physical system and the theoretical tools
employed to study and analyze its properties.
In Sec.~\ref{RESULTS} we disseminate the results
for the ground state (Subsec.~\ref{STATICS}), 
the dynamics (Subsec.~\ref{DYNAMICS}),
and for systems made of larger numbers of particles (Subsec.~\ref{LARGE_SYS}).
Concluding remarks are given in Sec.~\ref{CONCLUSIONS}.
Finally, Appendix \ref{MOMENTUM} discusses the behavior of the 
many-particle momentum variance.

\section{System and Methodology}\label{THEORY}

We consider the many-particle Schr\"odinger equation in two spatial dimensions
for the ground state,
$\hat H(\r_1,\ldots,\r_N)\Phi(\r_1,\ldots,\r_N)=E\Phi(\r_1,\ldots,\r_N)$,
and out-of-equilibrium dynamics,
$\hat H(\r_1,\ldots,\r_N)\Psi(\r_1,\ldots,\r_N;t)=i\frac{\partial\Psi(\r_1,\ldots,\r_N;t)}{\partial t}$.
The many-body Hamiltonian is
$\hat H(\r_1,\ldots,\r_N) = 
\sum_{j=1}^N \left[ -\frac{1}{2}\frac{\partial^2}{\partial \r_j^2} + \hat V(\r_j) \right] + \sum_{j<k} \lambda_0 \hat W(\r_j-\r_k)$.
From the wavefunction $\Phi(\r_1,\ldots,\r_N)$, assumed to be normalized to $1$,
the reduced one-particle density matrix
$\rho(\r,\r')=N\int d\r_2 \cdots d\r_N\Phi(\r,\r_2,\ldots,\r_N)\Psi^\ast(\r',\r_2,\ldots,\r_N)=
\sum_j n_j \phi_j(\r) \phi^\ast_j(\r')$
and reduced two-particle density matrix
$\rho(\r_1,\r_2,\r'_1,\r'_2)=N(N-1)\int d\r_3 \cdots d\r_N\Psi(\r_1,\r_2,\r_3,\ldots,\r_N)\break\hfill\Psi^\ast(\r'_1,\r'_2,\r_3,\ldots,\r_N)$
are defined, and analogously for the dynamics as a function of $t$.
The one-particle density, or, simply, the density is the diagonal $\rho(\r)=\rho(\r,\r')$,
and the diagonalization of $\rho(\r,\r')$ defines
the natural orbitals $\{\phi_j(\r)\}$ and occupation numbers $\{n_j\}$.
From the latter, the number of depleted particles is given by $\sum_{j>1} n_j =N-n_1$. 

The trap potential is given by $\hat V(\r)=0.05\r^4+V_0e^{-\frac{\r^2}{2}}$,
with a barrier of heights $V_0=5$, $10$, $50$, and $100$ throughout this work.
The trap is radial and anharmonic, and the central barrier is a Gaussian.
The shape of the trapping potential is that of an annulus.
As the height $V_0$ of the barrier is increased, 
the radius of the annulus increases and its thickness decreases,
see Fig.~\ref{f1} and discussion below.
The interaction between bosons is repulsive and taken to be 
$\lambda_0W(\r-\r')=\lambda_0e^{-\frac{(\r-\r')^2}{2}}$, i.e., $\lambda_0>0$ throughout this work.
We focus in the present work on weak repulsion
which leads to a small amount of depletion of the bosons, see Fig.~\ref{f2} and discussion below.
The shape and range of the interaction do not
have a qualitative influence on the physics to be described hereafter.

We employ the multiconfigurational time-dependent Hartree 
for bosons (MCTDHB) method \cite{MCTDHB1,MCTDHB2}.
MCTDHB uses for the wavefunction
a variationally optimal ansatz 
which is a linear-combination
of all permanents generated  
by distributing the $N$ bosons over $M$ time-adaptive orbitals.
As the number of such orbitals is increased,
convergence of quantities with $M$ in obtained.
These concepts have been well documented and discussed in the literature,
and we therefore keep the presentation here concise.
For applications, benchmarks, and extensions see \cite{MCTDH_BB,BJJ,MCTDHB_OCT,
Benchmarks,ML1,ML2,MCTDHB_3D_stat,MCTDHB_3D_dyn,Breaking,Uwe,2D_Tun,
Axel_MCTDHF_HIM,MCTDHB_spin,Kaspar_n,ML3,Higher_DIM,Cami_NJP,Axel_ar,Axel_Cavity,Dimensional_Cross,2D_Dark,Peter_Mix1,Peter_Mix2,Cami_JPB,Axel_2018,Alexej_2018}. 
We use the numerical implementation \cite{PACK1,PACK2} in imaginary time for the ground state \cite{MCHB}
and real time for the dynamics.
Finally, it should be mentioned that MCTDHB is the bosonic variant of the highly-efficient distinguishable-particle
multiconfigurational time-dependent Hartree method amply used in molecular physics 
\cite{cpl1990,jcp1992,review_Dieter,book_MCTDH,ML-MCTDH,ML-Manthe,ML-Oriol}.

For the numerical solution we use a grid of $64^2$ points in a box of size $[-8,8) \times [-8,-8)$ with periodic boundary conditions.
Convergence of the results with the number of grid points has been checked using a grid of $128^2$ points.
The dense grid of $256^2$ points used in the computations of the densities in Fig.~\ref{f1} is for the 
accurate determination of the radius of the density at its peak.

\section{Results and discussion}\label{RESULTS}

We first study static and dynamic properties of a finite system 
made of $N=10$ bosons in Subsecs.~\ref{STATICS} and \ref{DYNAMICS}, respectively,
on the basis of which we proceed to investigate larger systems in Subsec.~\ref{LARGE_SYS}.
We point out that in all studied cases the bosons are essentially or mostly condensed. 

\subsection{Statics}\label{STATICS}

\begin{figure}[!]
\begin{center}
\includegraphics[width=0.64\columnwidth,angle=-90]{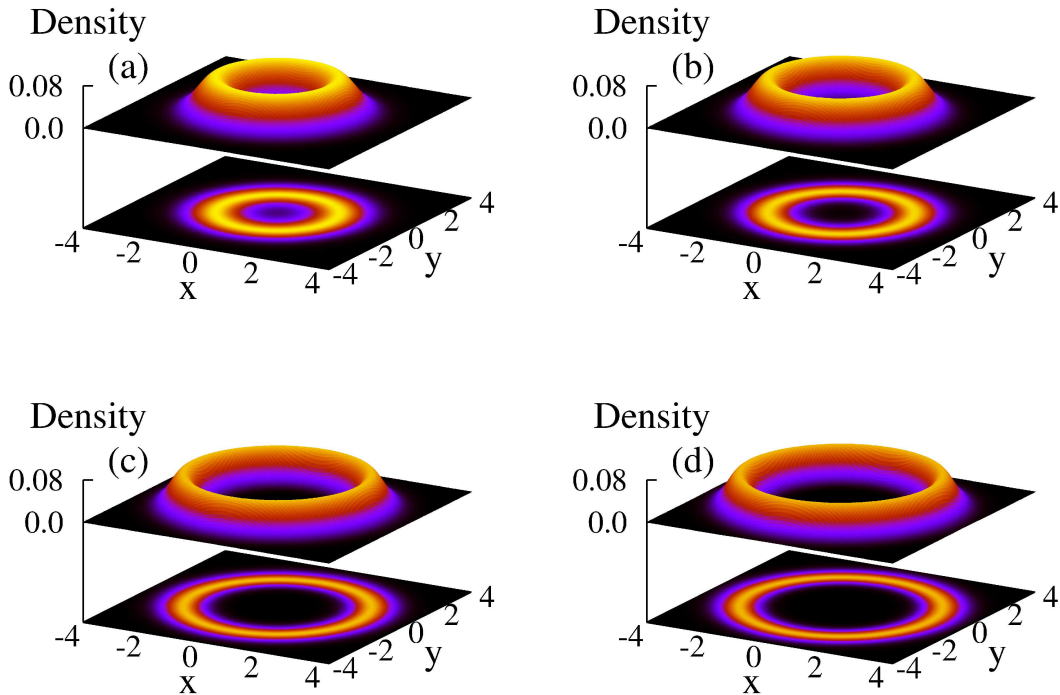}
\end{center}
\caption{Ground-state density per particle, $\frac{\rho(\r)}{N}$,
of $N=10$ bosons in a two-dimensional annulus of
barrier heights (a) $V_0=5$,
(b) $V_0=10$,
(c) $V_0=50$,
and
(d) $V_0=100$.
The interaction strength is $\lambda_0=0.02$ and
the number of self-consistent orbitals used is $M=10$.
The radius of the density at its maximal value, $R$, is determined numerically as
(a) $R=1.75$, (b) $R=2.06$, (c) $R=2.62$, and (d) $R=2.87$.
See the text for more details.
The quantities shown are dimensionless.}
\label{f1}
\end{figure}

Fig.~\ref{f1} shows the ground-state
density per particle, $\frac{\rho(\r)}{N}$, of $N=10$ bosons for the four annular 
traps of growing size generated from the barrier heights $V_0=5$, $10$, $50$, and $100$. 
The radius of the density at its maximal value, $R$, is determined numerically
using a computation with a resolution of $256^2$ grid points as
(a) $R=1.75(0)$, (b) $R=2.06(2)$, (c) $R=2.62(5)$, and (d) $R=2.87(5)$.

\begin{figure}[!]
\begin{center}
\hglue -1.25 truecm
\includegraphics[width=0.2630\columnwidth,angle=-90]{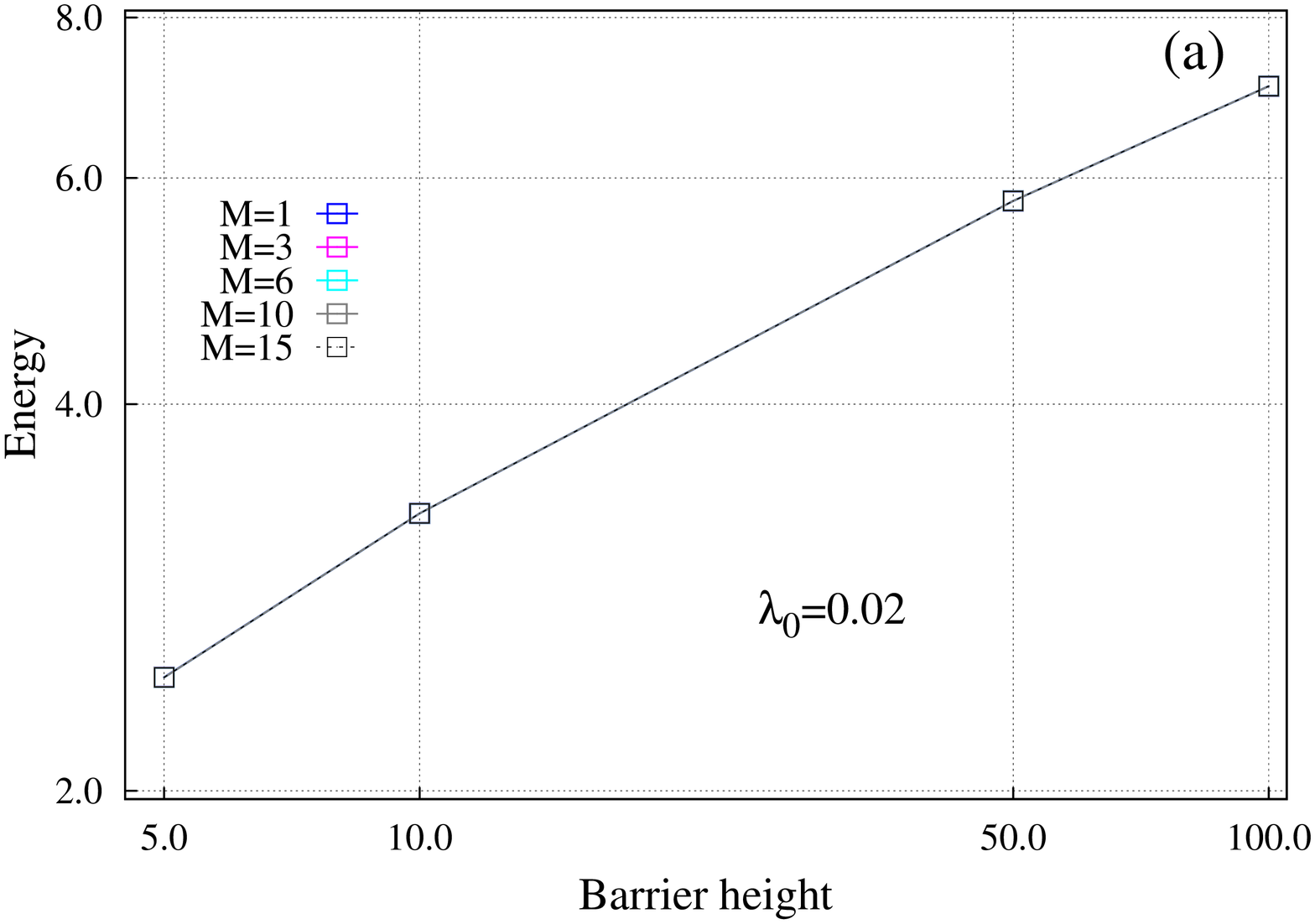}\hglue -0.25 truecm
\includegraphics[width=0.2630\columnwidth,angle=-90]{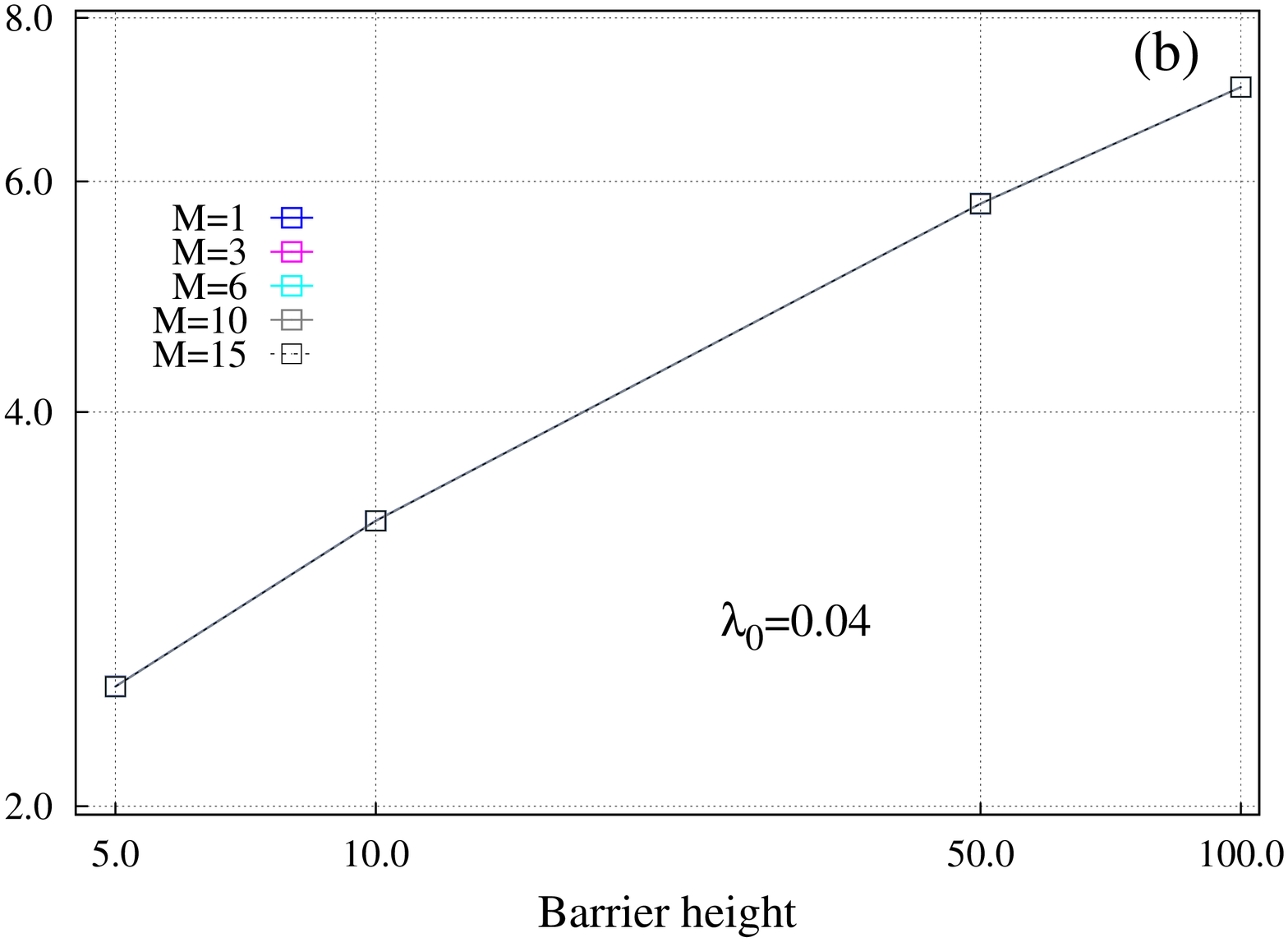}\hglue -0.25 truecm
\includegraphics[width=0.2630\columnwidth,angle=-90]{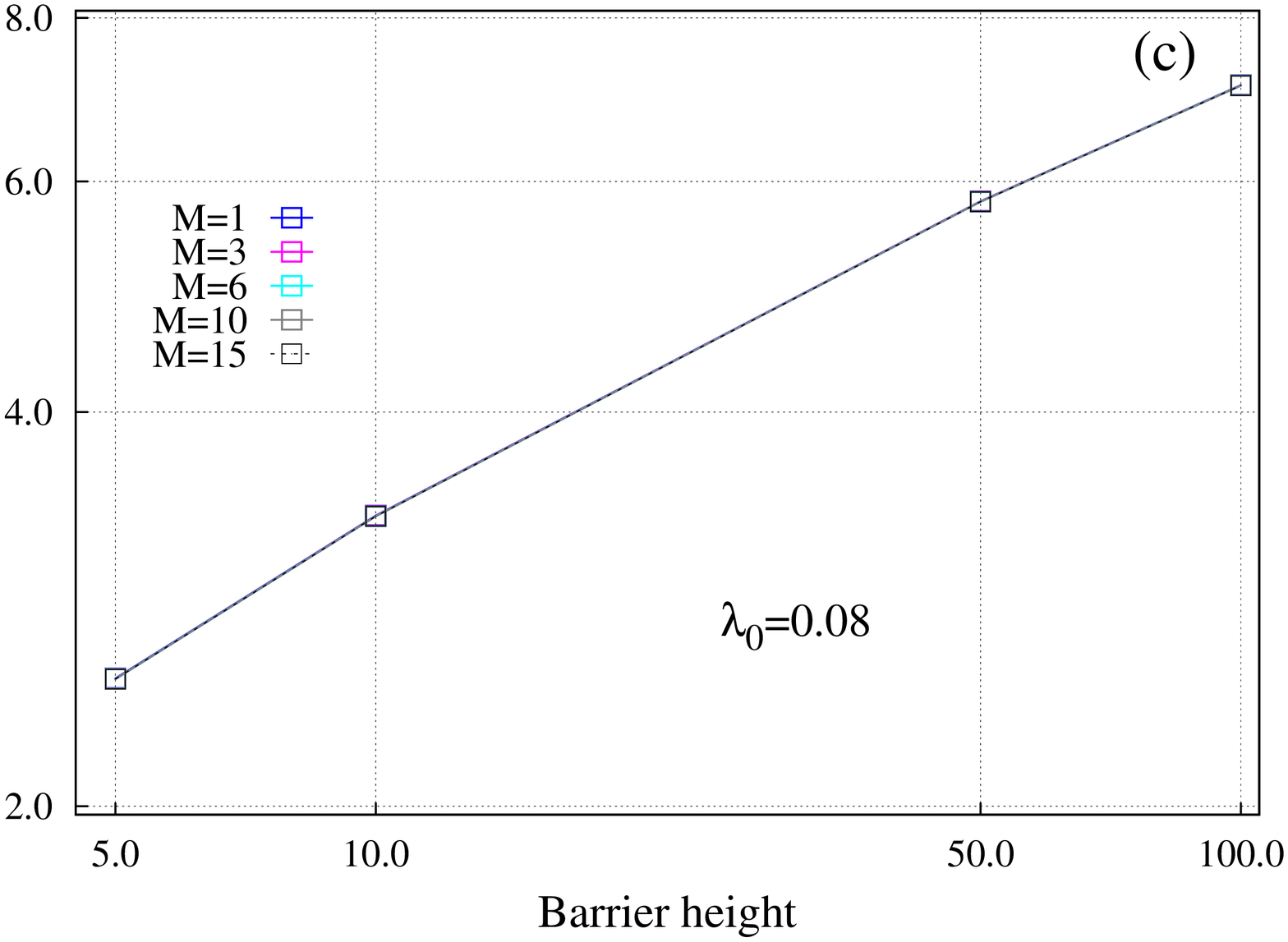}
\vglue 0.25 truecm
\hglue -1.25 truecm
\includegraphics[width=0.2630\columnwidth,angle=-90]{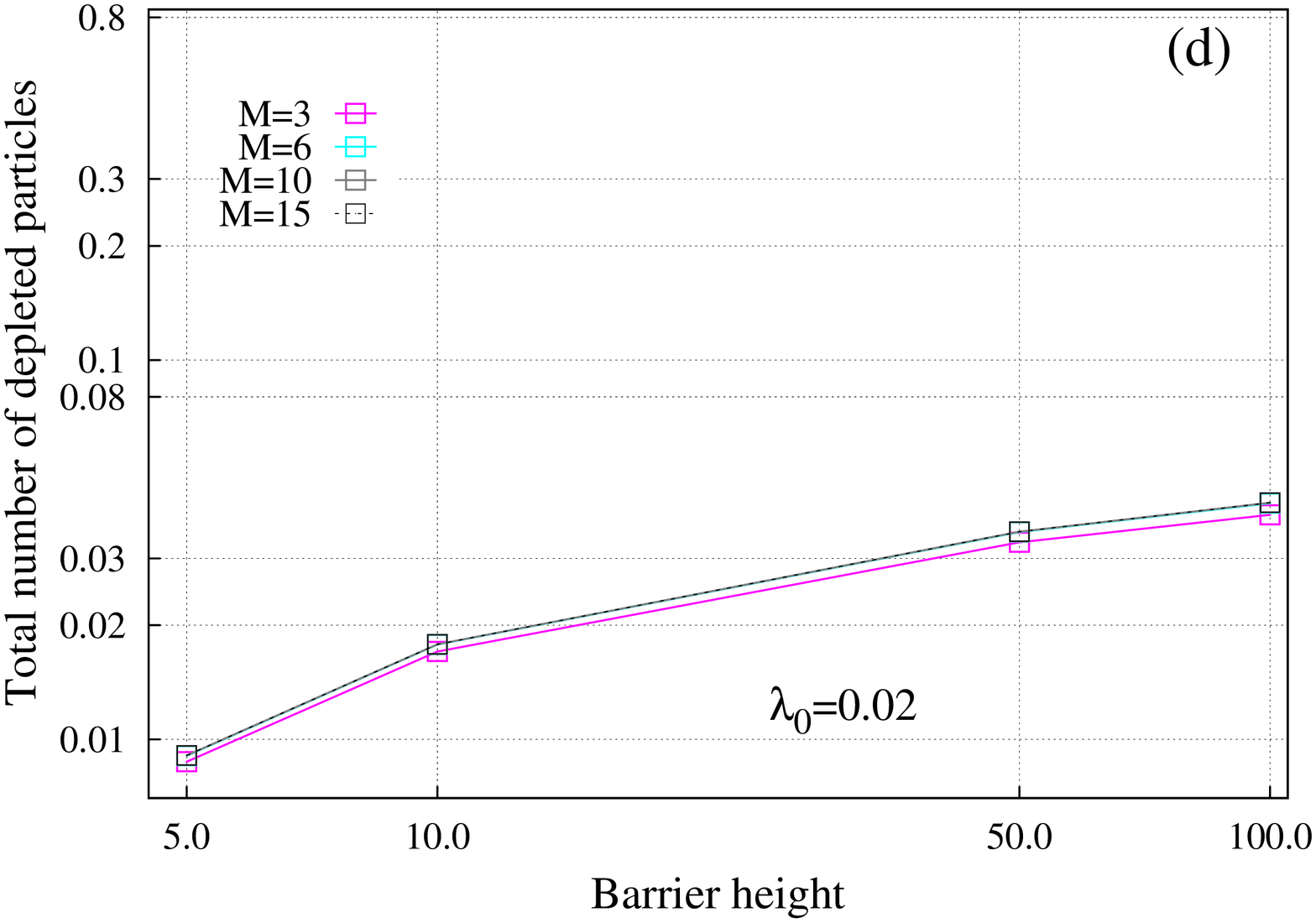}\hglue -0.25 truecm
\includegraphics[width=0.2630\columnwidth,angle=-90]{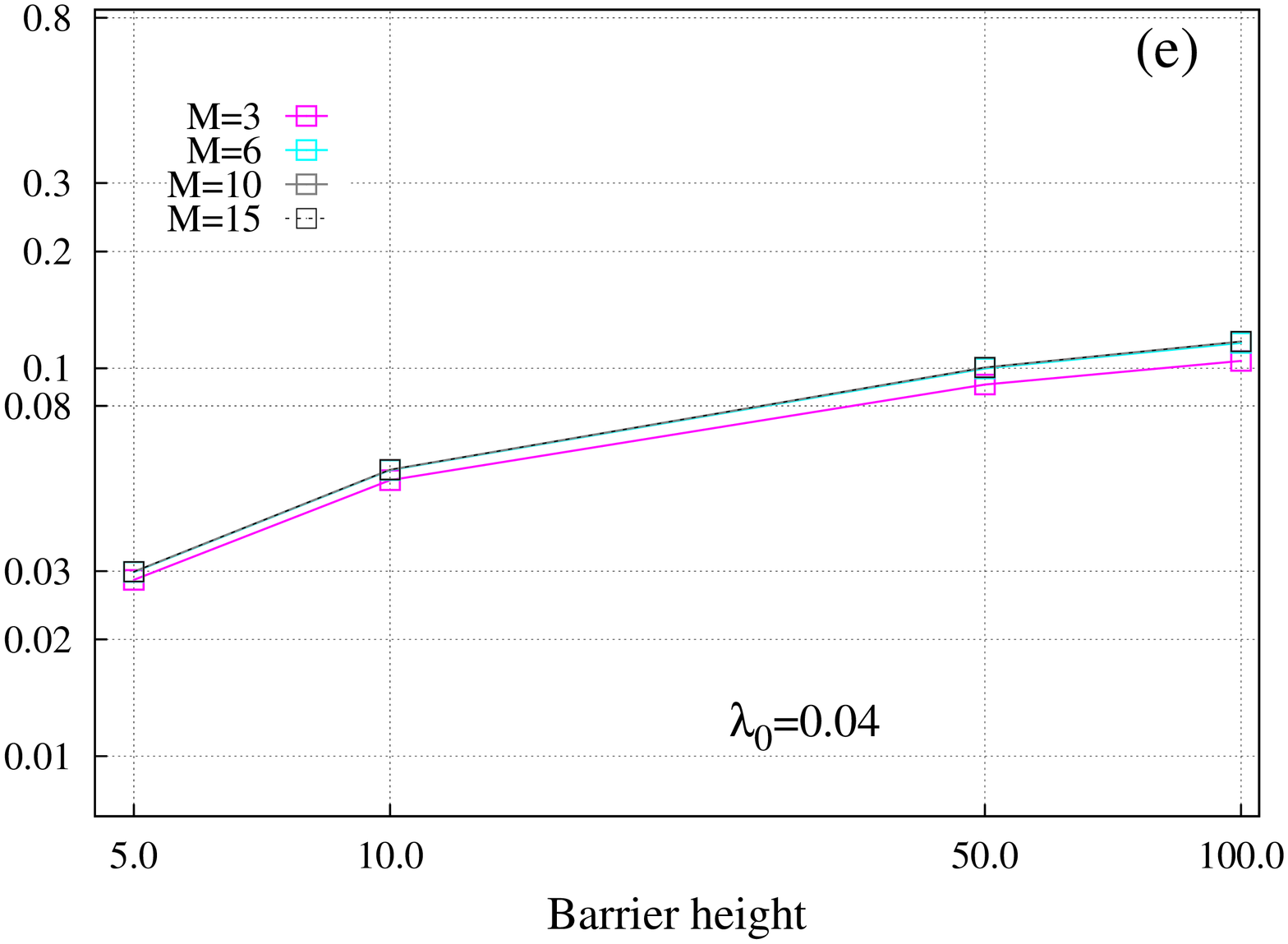}\hglue -0.25 truecm
\includegraphics[width=0.2630\columnwidth,angle=-90]{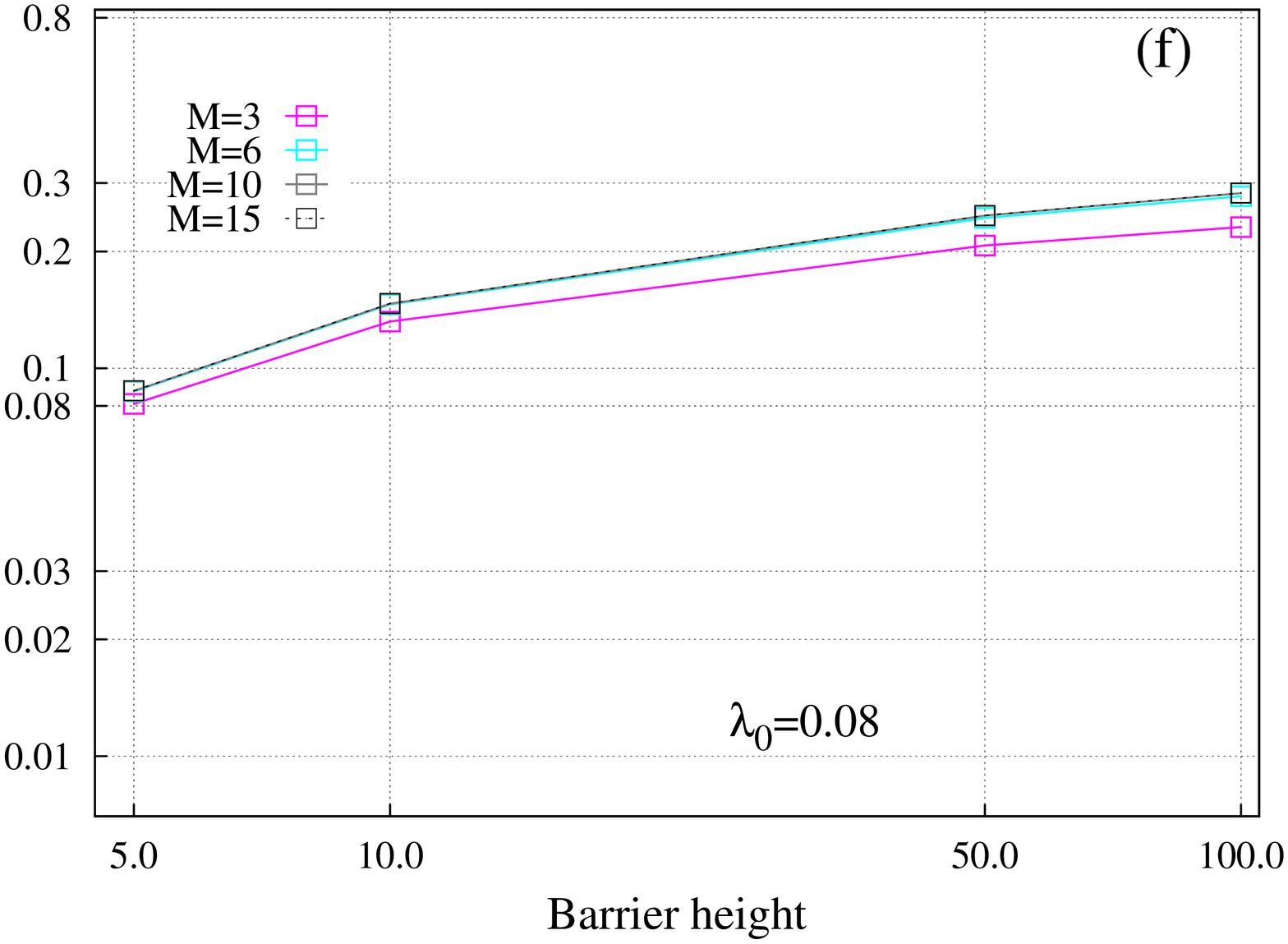}
\vglue 0.25 truecm
\hglue -1.25 truecm
\includegraphics[width=0.2630\columnwidth,angle=-90]{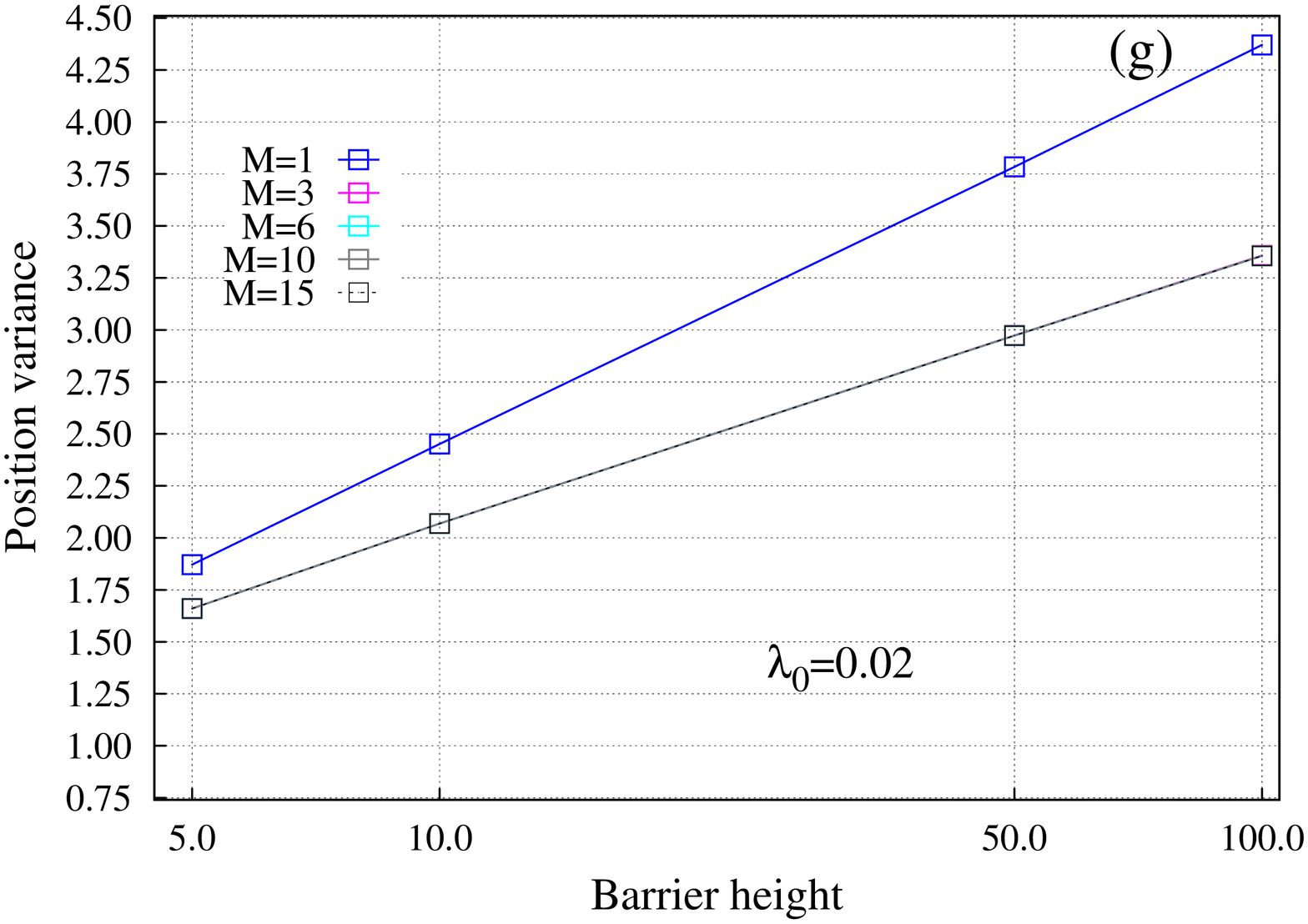}\hglue -0.25 truecm
\includegraphics[width=0.2630\columnwidth,angle=-90]{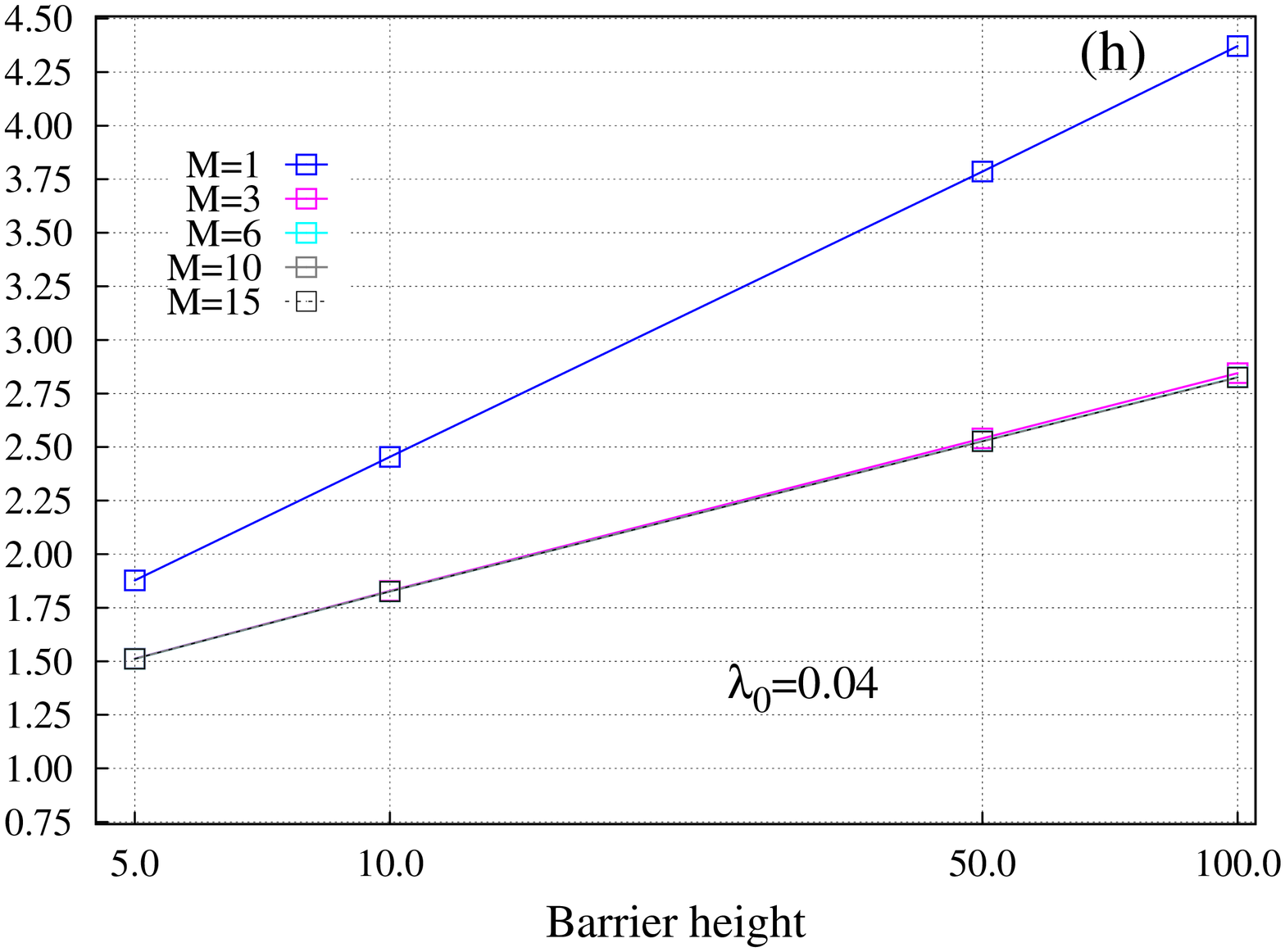}\hglue -0.25 truecm
\includegraphics[width=0.2630\columnwidth,angle=-90]{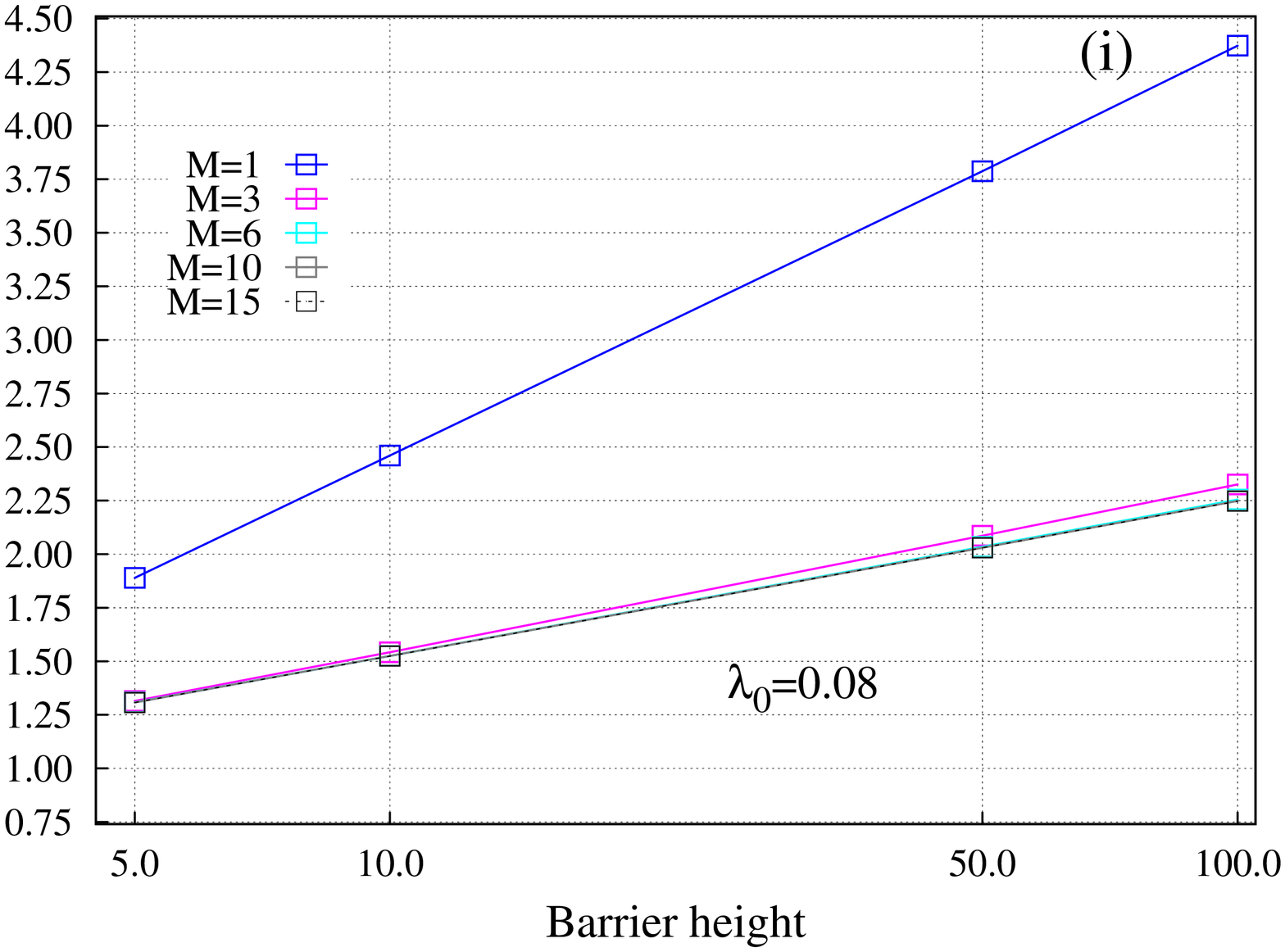}
\end{center}
\vglue 0.75 truecm
\caption{Ground-state properties as a function of the barrier height $V_0$ and interaction strength $\lambda_0$. 
Energy per particle, $\frac{E}{N}$ [top row, panels (a), (b), and (c)],
total number of depleted particles, $N-n_1$ [middle row, panels (d), (e), and (f)],
and many-particle position variance per particle, $\frac{1}{N}\Delta^2_{\hat X}$ [lower row, panels (g), (h), and (i)].
The number of bosons is $N=10$.
The barrier heights are $V_0=5$, $10$, $50$, and $100$.
The interaction strengths are $\lambda_0=0.02$ [left column, panels (a), (d), and (g)],
$\lambda_0=0.04$ [middle column, panels (b), (e), and (h)],
and $\lambda_0=0.08$ [right column, panels (c), (f), and (i)].
Actual data are marked by symbols, the continuous curves are to guide the eye only.
See the text for more details.
The quantities shown are dimensionless.}
\label{f2}
\end{figure}

Fig.~\ref{f2} depicts the energy per particle $\frac{E}{N}$,
total number of depleted particles $N-n_1$, 
and the many-body position variance per particle,
$\frac{1}{N} \Delta^2_{\hat X} = \int d\r \frac{\rho(\r)}{N}x^2 - N\left[\int d\r \frac{\rho(\r)}{N}x\right]^2
+ \int d\r_1 d\r_2 \frac{\rho^{(2)}(\r_1,\r_2,\r_1,\r_2)}{N}x_1x_2$,
of the $N=10$ bosons
for the three interaction strengths $\lambda_0=0.02$, $0.04$, and $0.08$,
for the four barrier heights $V_0=5$, $10$, $50$, and $100$,
(i.e., the four annular traps with sizes $R$).

The energy plotted in the upper row of Fig.~\ref{f2} increases with the annulus size $R$.
This is because the one-body potential of the barrier which pushes up the baseline of the energy as $V_0$ is increased.
Otherwise, the energy would have decreased for annuli of growing radius.
Importantly, for a given annulus the energy has little dependence on 
$\lambda_0$ for such weak interaction strengths.
Furthermore, $\frac{E}{N}$ at the mean-field level of theory, i.e., $M=1$ self-consistent orbitals,
and at the many-body level of theory, i.e., $M=3$, $6$, $10$, and $15$ self-consistent orbitals,
are essentially the same,
indicating that the systems are well within their mean-field regime.
For instance, for the smallest $V_0=5$ annulus and weakest interaction $\lambda_0=0.02$,
$\frac{E}{N}=2.45107$ ($M=1$) and $\frac{E}{N}=2.45085$ ($M=10$),
and for the largest $V_0=100$ annulus and strongest interaction $\lambda_0=0.08$,
$\frac{E}{N}=7.10565$ ($M=1$) and $\frac{E}{N}=7.10147$ ($M=10$).

The total number of depleted particles depicted in the middle row of Fig.~\ref{f2}
increases with the annulus size for a given interaction strength,
indicating that the gap between the ground state and first excited one-particle states decreases with $R$.
Of course, the depletion increases
with $\lambda_0$ for a given annulus.
Overall, the depletion out of $N=10$ bosons is small,
ranging from less than $\frac{1}{100}$-th of a particle 
(i.e., $0.1\%$) for the smallest annulus and weakest interaction,
to less than $0.3$ of a particle (i.e., $3\%$) for the largest annulus and strongest interaction.

The many-particle position variance per particle, $\frac{1}{N} \Delta^2_{\hat X}$, is shown 
in the lower row of Fig.~\ref{f2} as a function of the barrier height
and for the three interaction strengths.
There are several features that immediately pop out.
First, there is a difference between the mean-field and many-body quantities.
The many-body variance is always smaller than the mean-field one,
as is expected from repulsive bosons \cite{var1}.
Second, the mean-field variance is almost independent of the interaction strength for a given annulus,
much like the energy discussed above.
On the other hand,
the many-body variance decreases with the repulsion strength for a given annulus,
in accordance with the increase of the depletion analyzed above.
Combining these three observations, we see that the difference between the many-body
and mean-field variances increases with the annulus size
and with the repulsion strength.

It is instructive to analyze the position variance at the mean-level of theory, and
discuss its connection with the size of the bosonic density.
Given the ground-state Gross-Pitaevskii solution $\phi_{GP}(\r)$,
one has $\frac{\rho(\r)}{N}=|\phi_{GP}(\r)|^2$
and $\frac{\rho^{(2)}(\r_1,\r_2,\r_1,\r_2)}{N}=(N-1)|\phi_{GP}(\r_1)|^2|\phi_{GP}(\r_2)|^2$,
leading to $\frac{1}{N} \Delta^2_{\hat X} = \int d\r |\phi_{GP}(\r)|^2 x^2 - \left[\int d\r |\phi_{GP}(\r)|^2 x\right]^2$.
Thus, at the mean-field level the position variance is determined only
by the shape of the density, $\rho(\r)=N|\phi_{GP}(\r)|^2$.
 
Let us consider then delta-function ring densities,
$\rho(\r)=\frac{N}{2\pi R}\delta(r-R)$,
where $R$ are the above determined radii.
This would give us an estimate for the mean-field position variance in the limit of a narrow annulus of radius $R$.
A straightforward calculation in polar coordinates, $x=r\cos(\varphi)$, 
gives $\frac{1}{N} \Delta^2_{\hat X} = 
\int \frac{1}{2\pi R}\delta(r-R) r^2 \cos^2(\varphi) r dr d\varphi = \frac{R^2}{2}$.
Plugging in the above radii, one finds (to two significant digits after the dot without rounding) 
$\frac{1}{N} \Delta^2_{\hat X}=1.53, 2.12, 3.44, 4.13$.
Comparing to the mean-field results (for $\lambda=0.02$), see Fig.~\ref{f2}g,
we have
$\frac{1}{N} \Delta^2_{\hat X}=1.87, 2.45, 3.78, 4.37$.
The numerical mean-field values are somewhat larger than the analytical analysis
where the respective differences are about the same and equal $\sim 0.3$.
The analytical analysis of a delta-function ring density underestimates the mean-field variance,
because the density in the annulus does have a finite radial width,
and radii larger than $R$ 
contribute to the variance integral more than radii smaller than $R$.

\subsection{Dynamics}\label{DYNAMICS}

\begin{figure}[!]
\begin{center}
\hglue -1.0 truecm
\includegraphics[width=0.345\columnwidth,angle=-90]{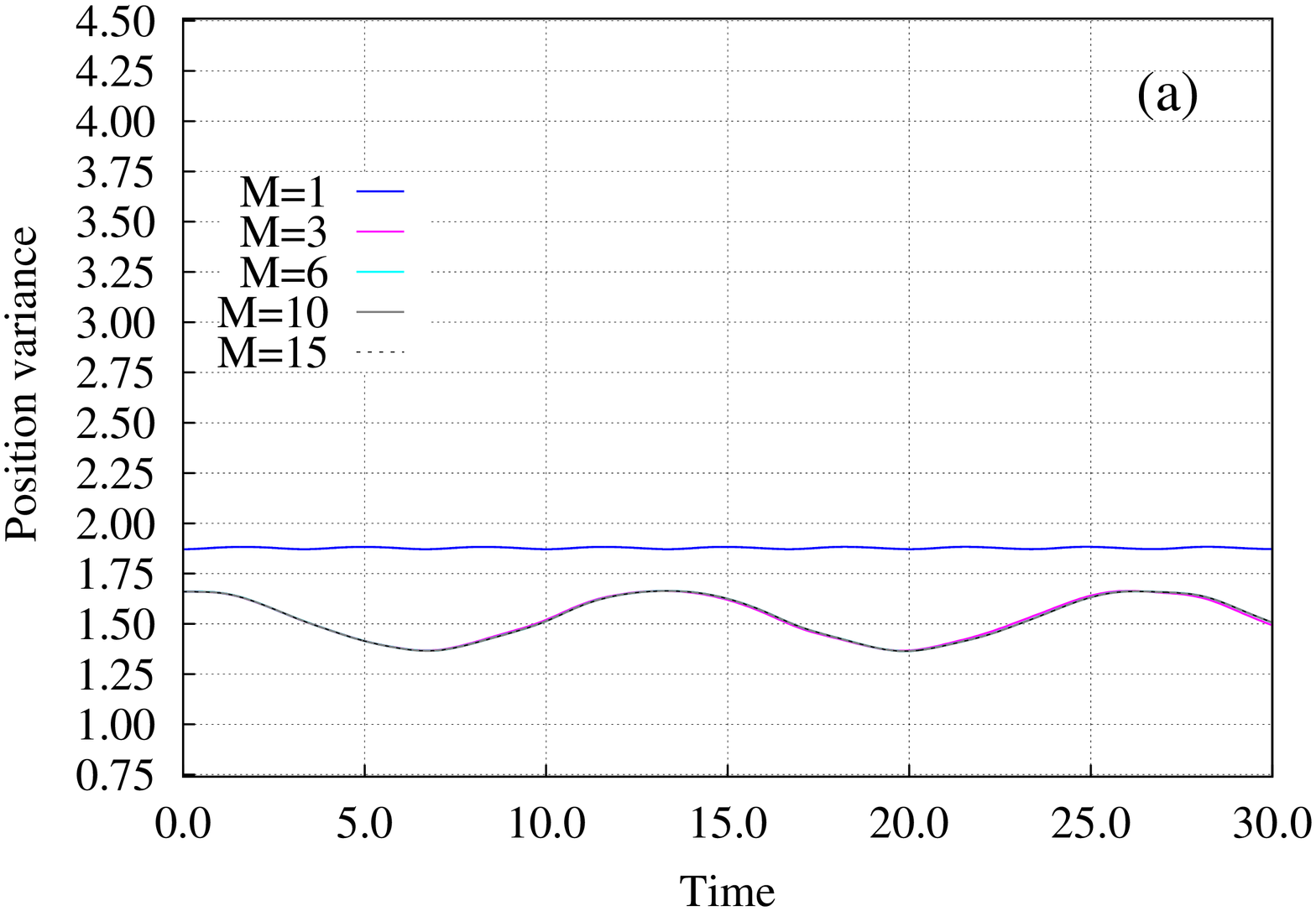}
\includegraphics[width=0.345\columnwidth,angle=-90]{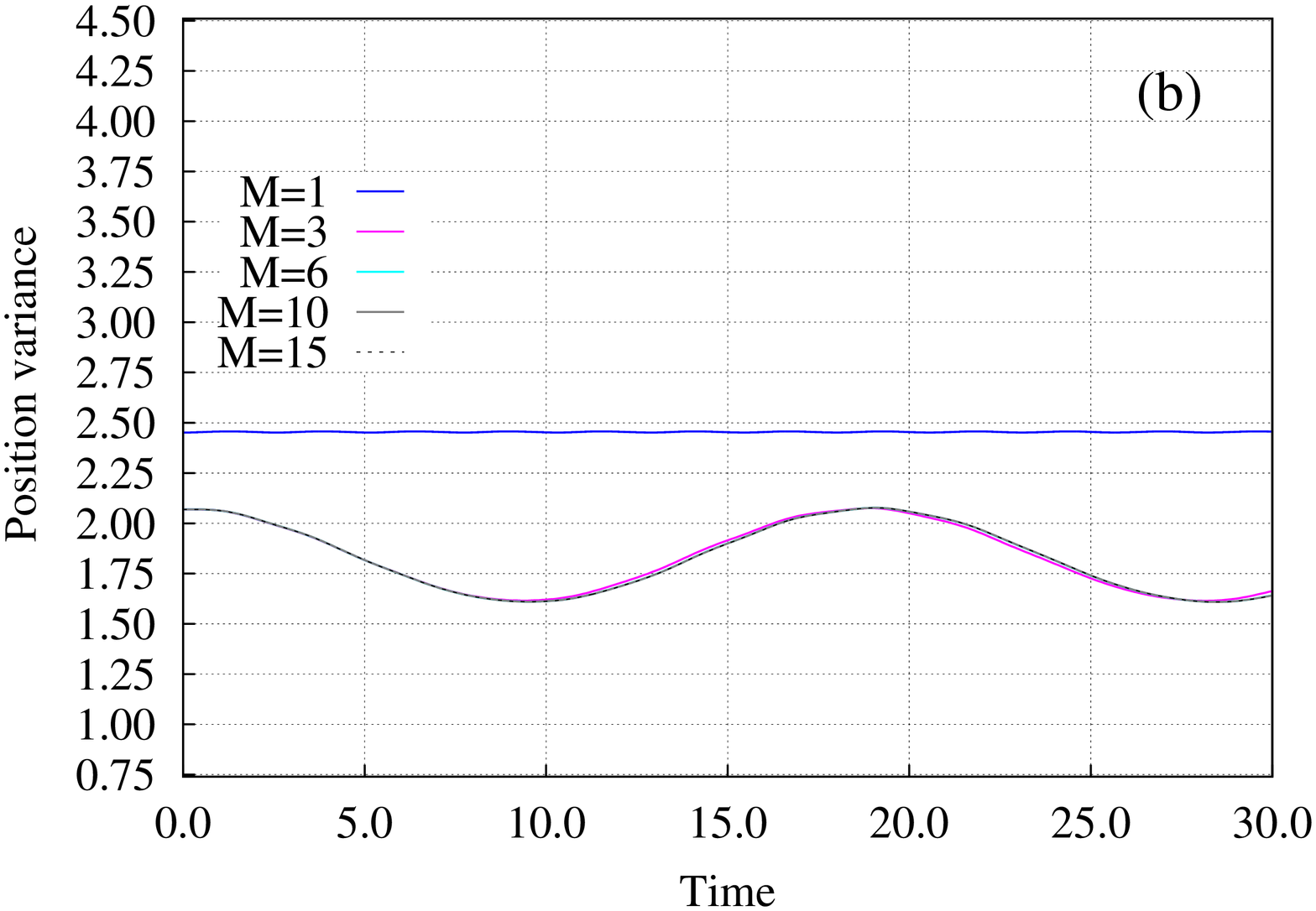}
\hglue -1.0 truecm
\includegraphics[width=0.345\columnwidth,angle=-90]{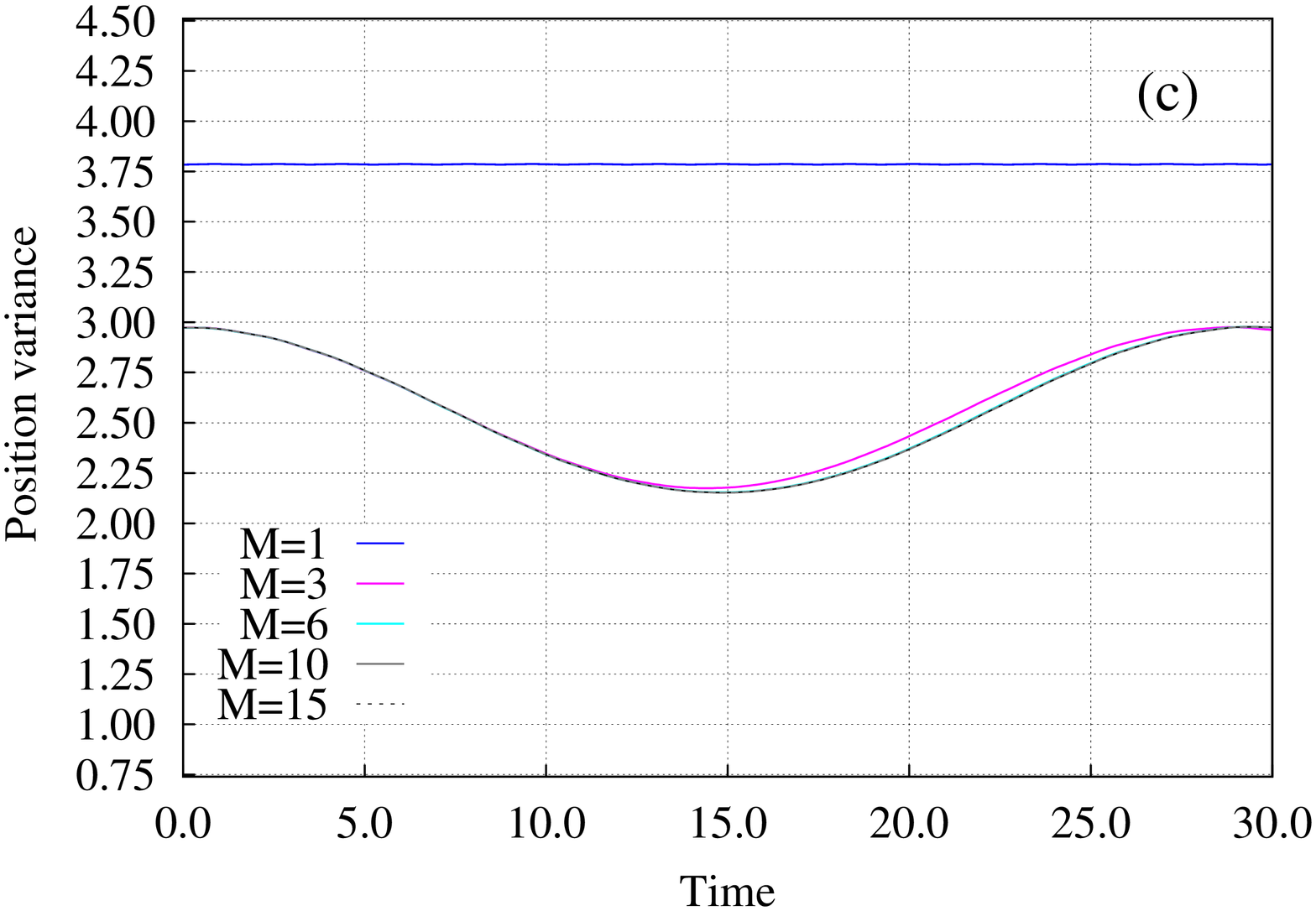}
\includegraphics[width=0.345\columnwidth,angle=-90]{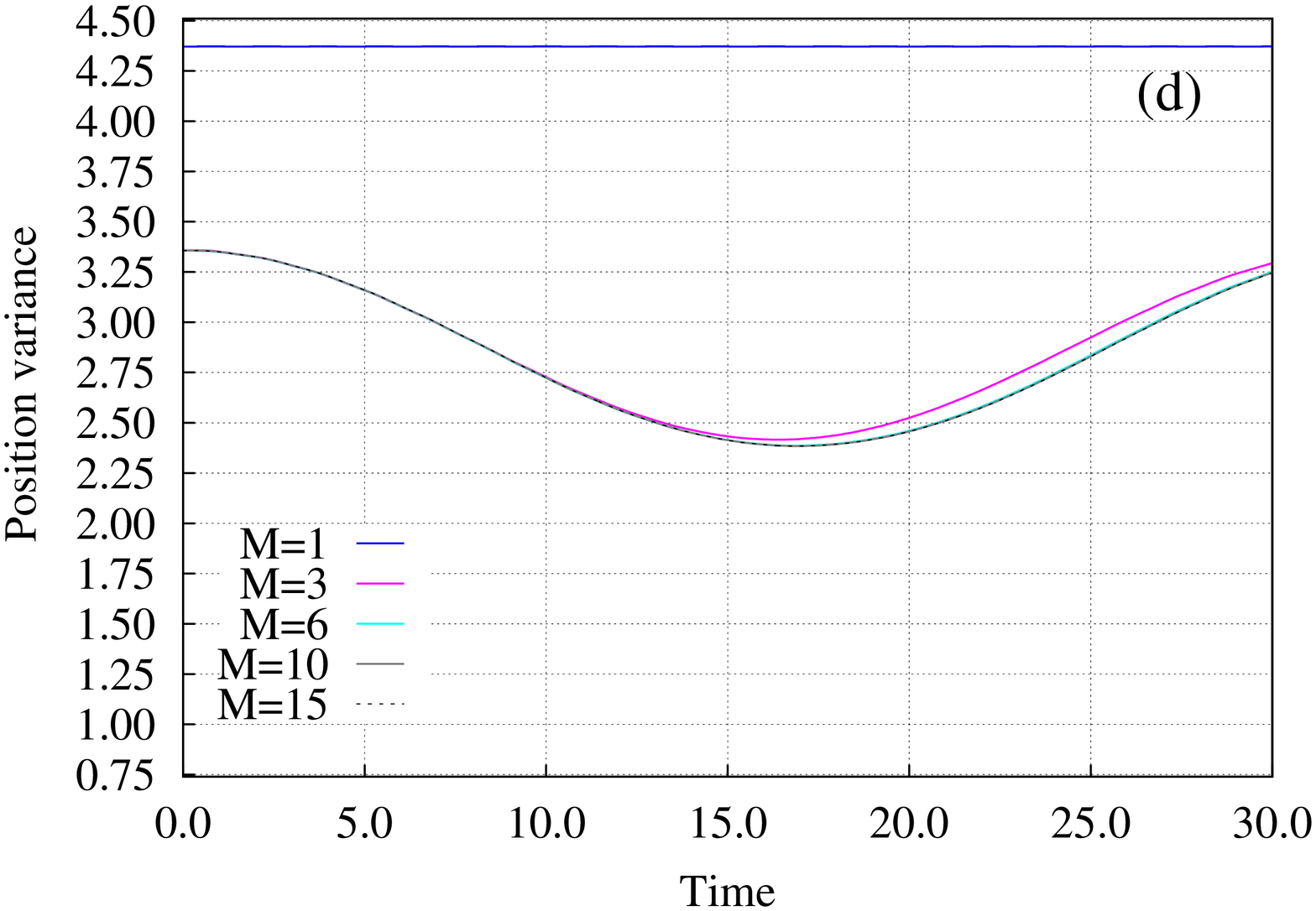}
\end{center}
\vglue 0.75 truecm
\caption{Variance breathing dynamics following a two-fold interaction quench.
The mean-field ($M=1$ time-adaptive orbitals) and many-body ($M=3$, $6$, $10$, and $15$ time-adaptive orbitals)
position variances per particle,
$\frac{1}{N}\Delta^2_{\hat X}(t)$,
of $N=10$ bosons in the annuli with barrier heights (a) $V_0=5$, (b) $V_0=10$, (c) $V_0=50$, and (d) $V_0=100$
following an interaction quench from $\lambda_0=0.02$ to $\lambda_0=0.04$.
The respective depletions are plotted in Fig.~\ref{f5}.
See the text for more details.
The quantities shown are dimensionless.}
\label{f3}
\end{figure}

\begin{figure}[!]
\begin{center}
\hglue -1.0 truecm
\includegraphics[width=0.345\columnwidth,angle=-90]{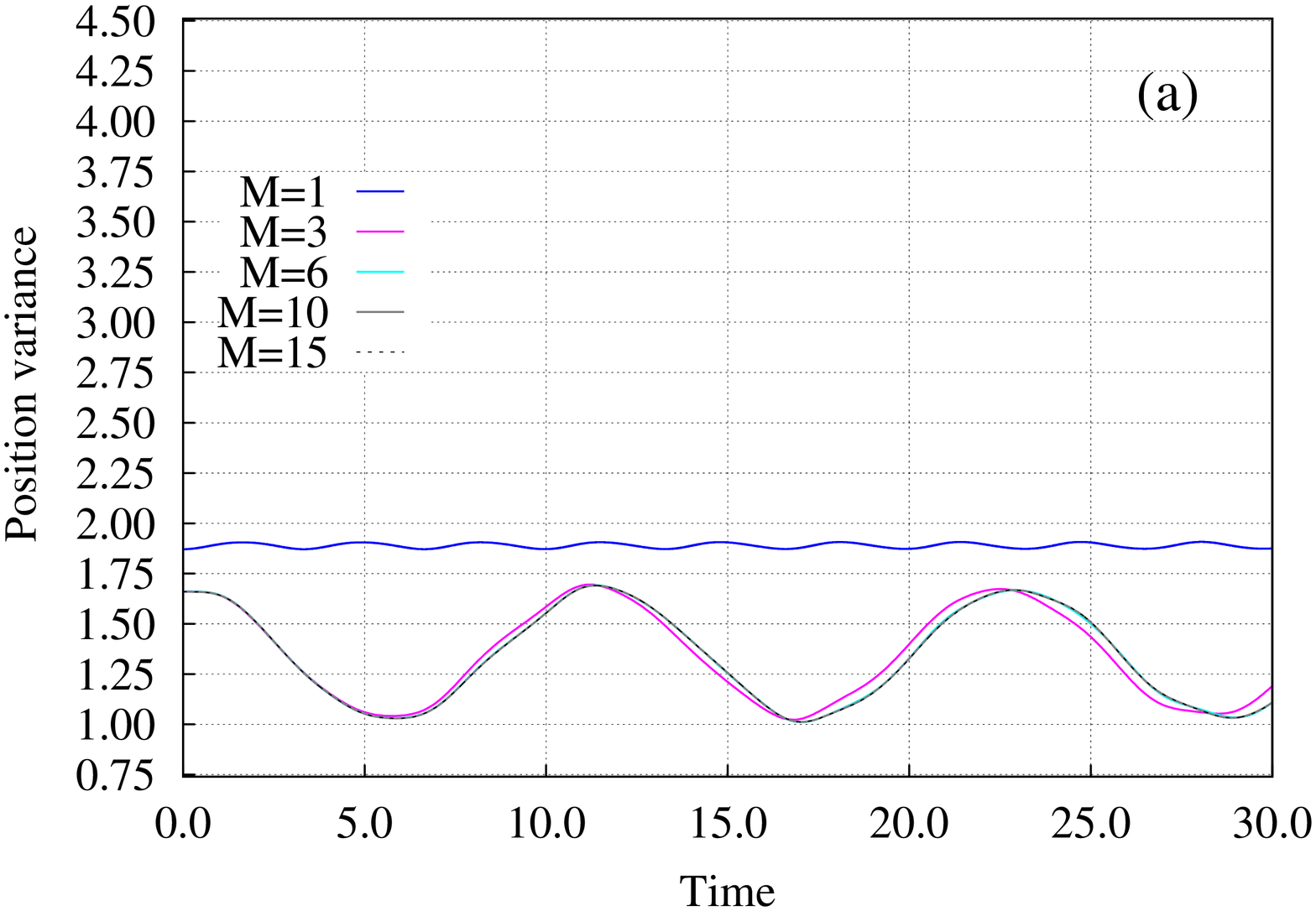}
\includegraphics[width=0.345\columnwidth,angle=-90]{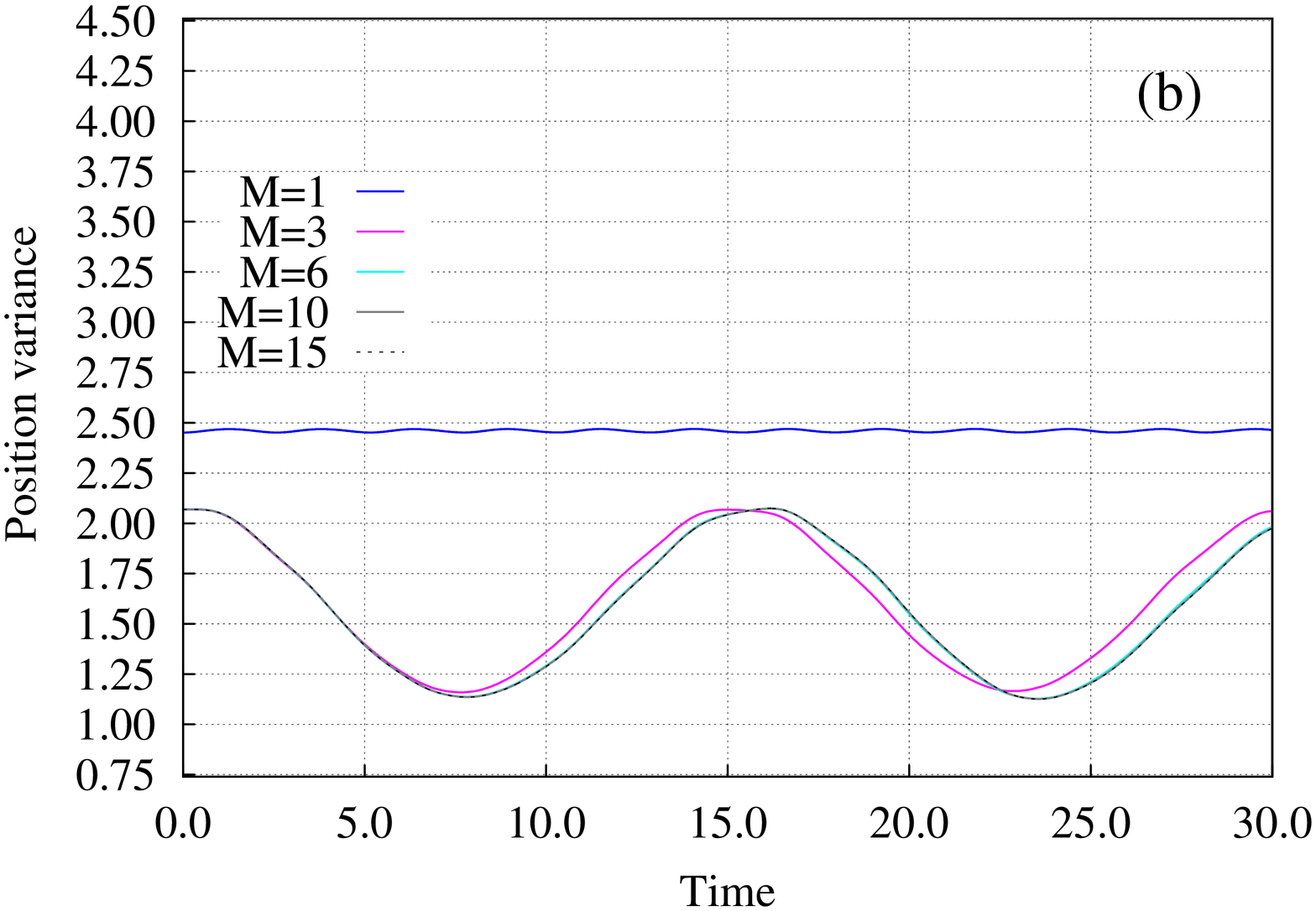}
\hglue -1.0 truecm
\includegraphics[width=0.345\columnwidth,angle=-90]{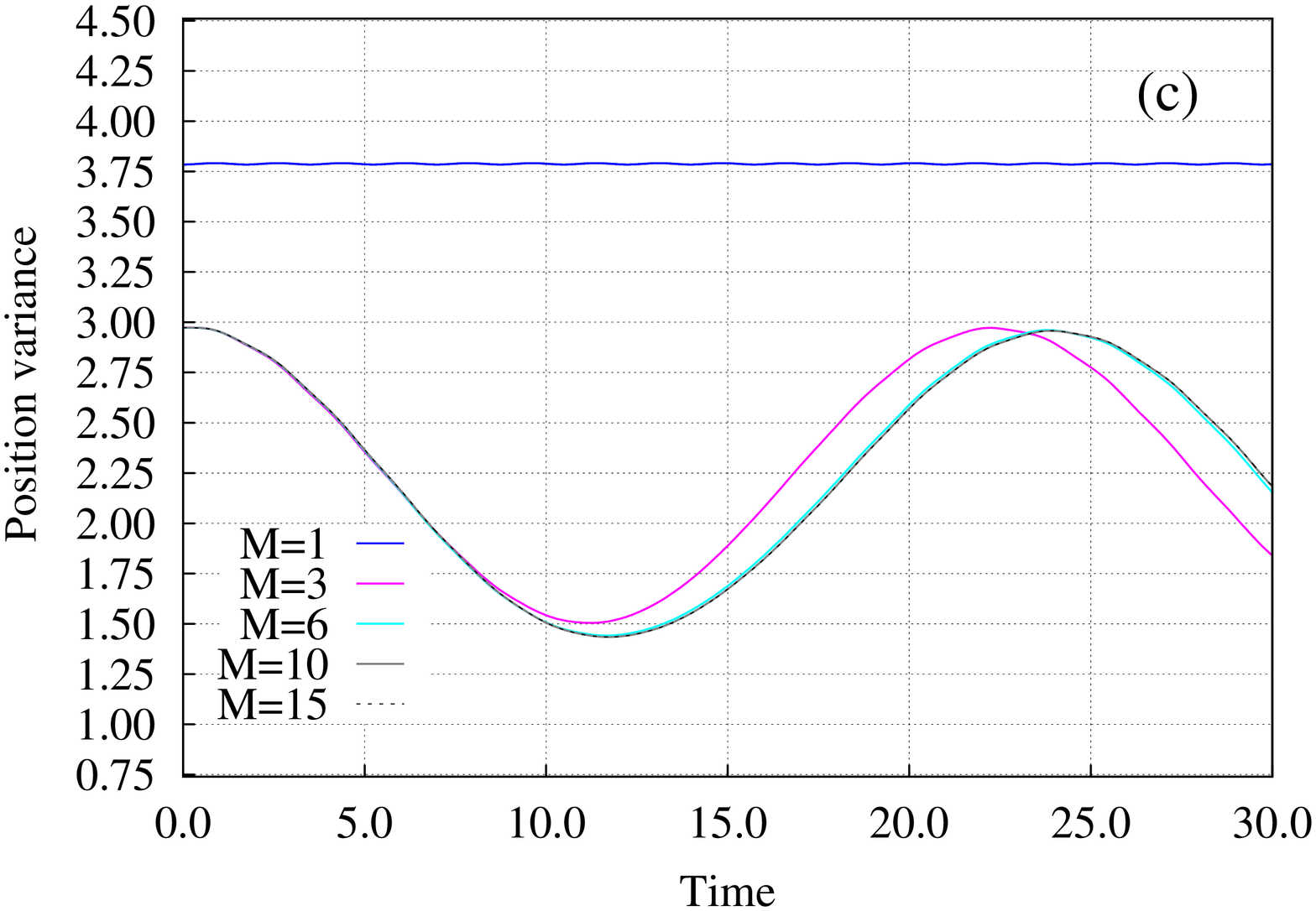}
\includegraphics[width=0.345\columnwidth,angle=-90]{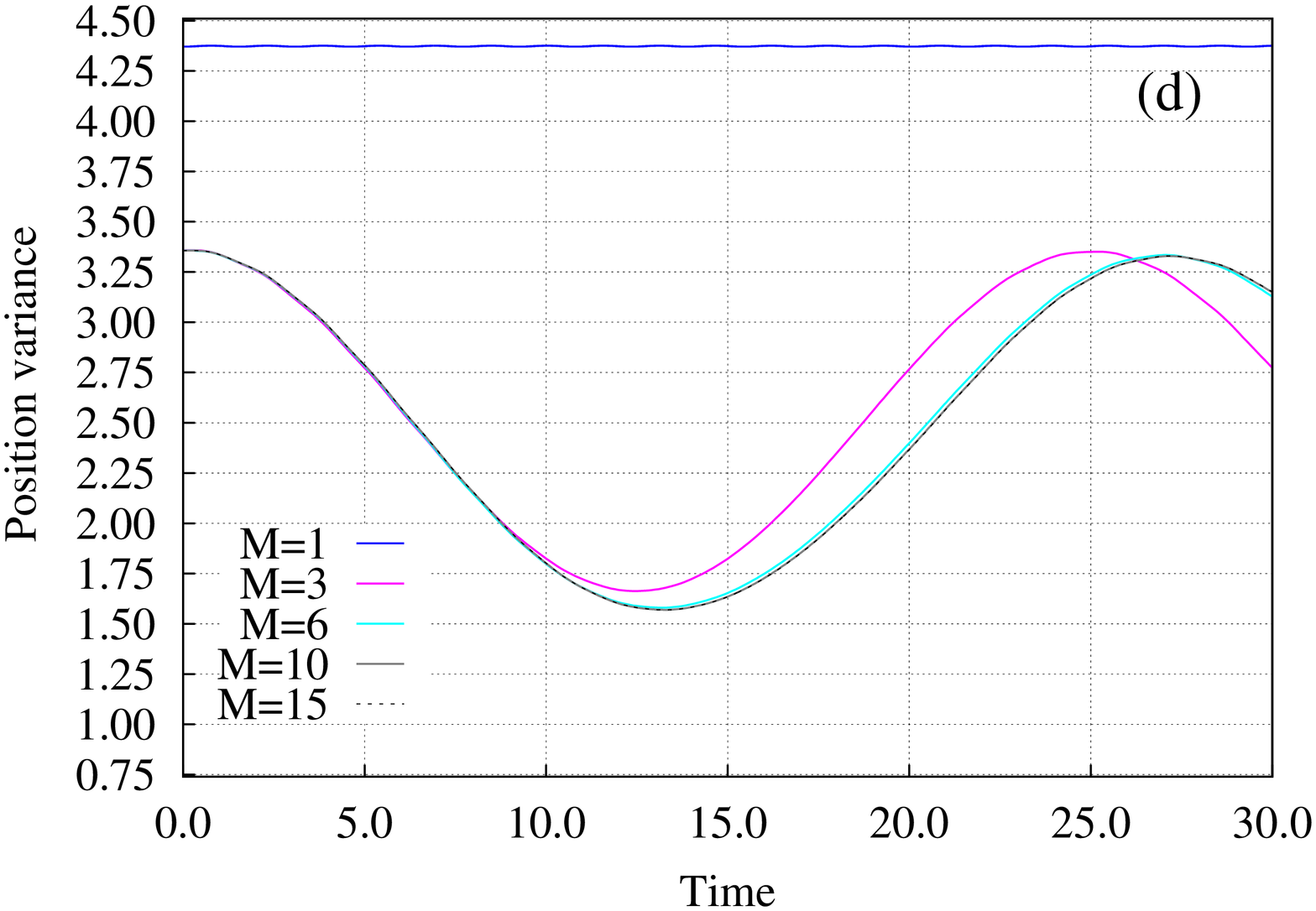}
\end{center}
\vglue 0.75 truecm
\caption{Variance breathing dynamics following a four-fold interaction quench.
The mean-field ($M=1$ time-adaptive orbitals) and many-body ($M=3$, $6$, $10$, and $15$ time-adaptive orbitals)
position variances per particle,
$\frac{1}{N}\Delta^2_{\hat X}(t)$,
of $N=10$ bosons in the annuli with barrier heights (a) $V_0=5$, (b) $V_0=10$, (c) $V_0=50$, and (d) $V_0=100$
following an interaction quench from $\lambda_0=0.02$ to $\lambda_0=0.08$.
The respective depletions are plotted in Fig.~\ref{f5}.
See the text for more details.
The quantities shown are dimensionless.}
\label{f4}
\end{figure}

The properties of the ground-state discussed above tell us that the energy weakly depends on the interaction, 
and that the mean-field variance which accounts for the shape of the density also weakly depends on the interaction.
On the other hand, the small amount of depletion pronouncedly depends on the interaction and this, in turn, leads
to large deviations of the many-body variance from the mean-field variance.
Recall that the center of mass is at the origin ($\langle \hat X \rangle = 0$).
The values of the many-body position variance become 
(much) smaller than the radius or size of the annulus.
At the many-body level, the fluctuations in the positions of the bosons
`live' in the wide regime inside the annulus,
not on the narrow annulus itself.
This suggests that, geometrically, the mean-field and many-body variances 
behave in a different manner and effectively exhibit different dimensionalities.
 
To get a more pronounced effect, and a microscopic explanation, we proceed to dynamics.
More precisely, we quench the interaction and monitor the time-evolution of the variance and the depletion.
The results and their analysis will tell as more on the dimensionality of the variance
and the participating excitations governing the effect.

\begin{figure}[!]
\begin{center}
\vglue -1.75 truecm
\hglue -1.0 truecm
\includegraphics[width=0.2930\columnwidth,angle=-90]{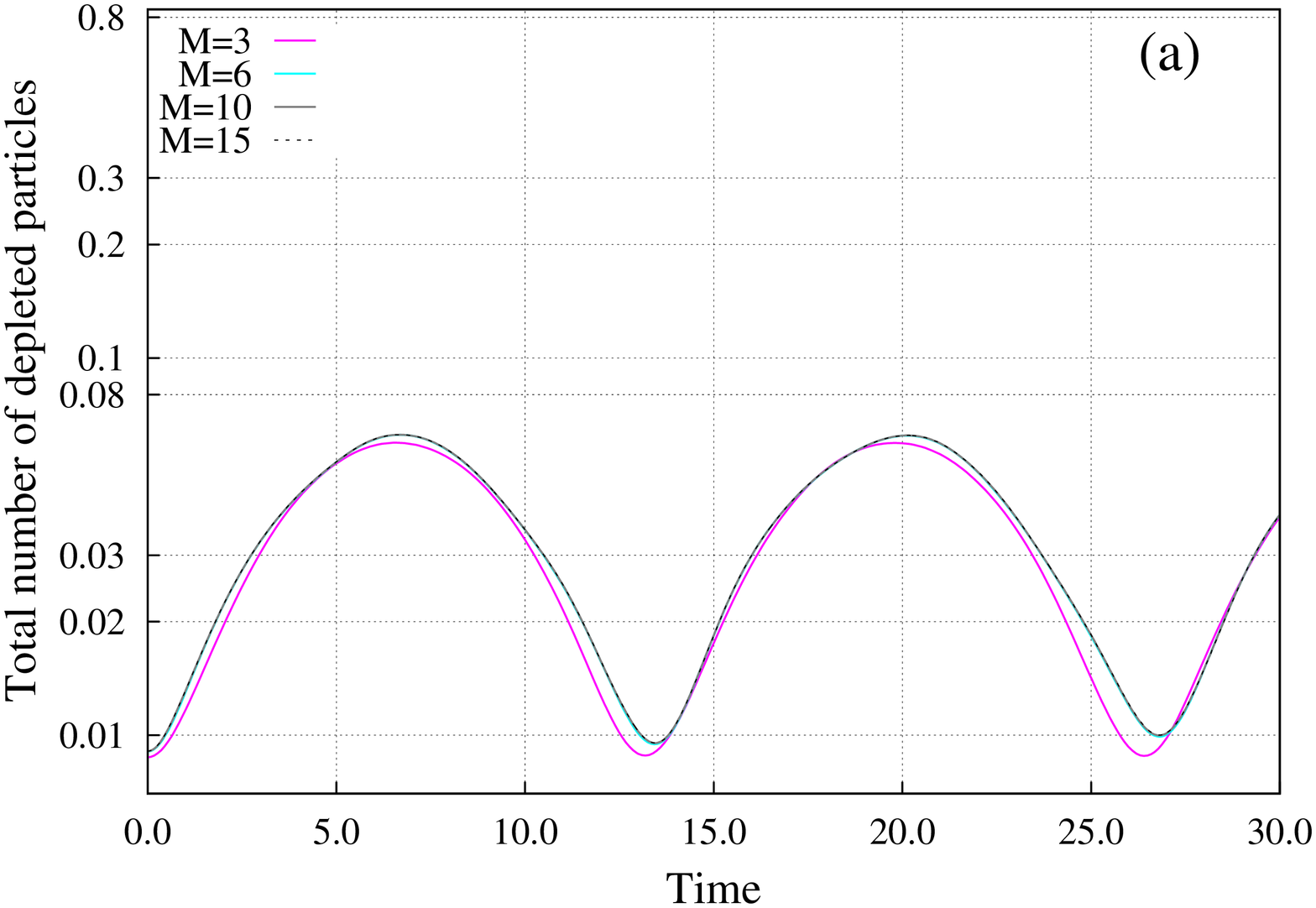}
\includegraphics[width=0.2930\columnwidth,angle=-90]{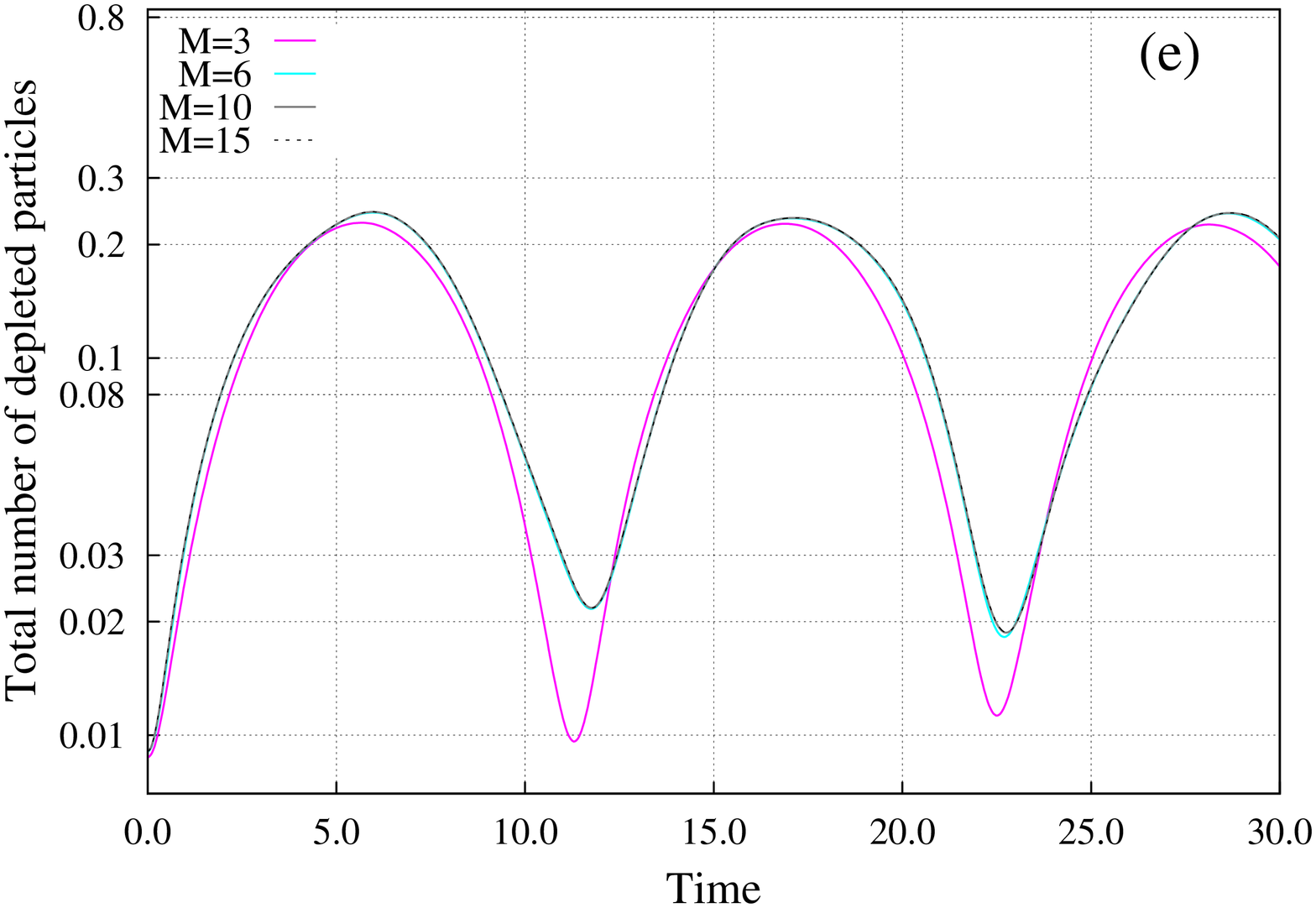}
\vglue 0.25 truecm
\hglue -1.0 truecm
\includegraphics[width=0.2930\columnwidth,angle=-90]{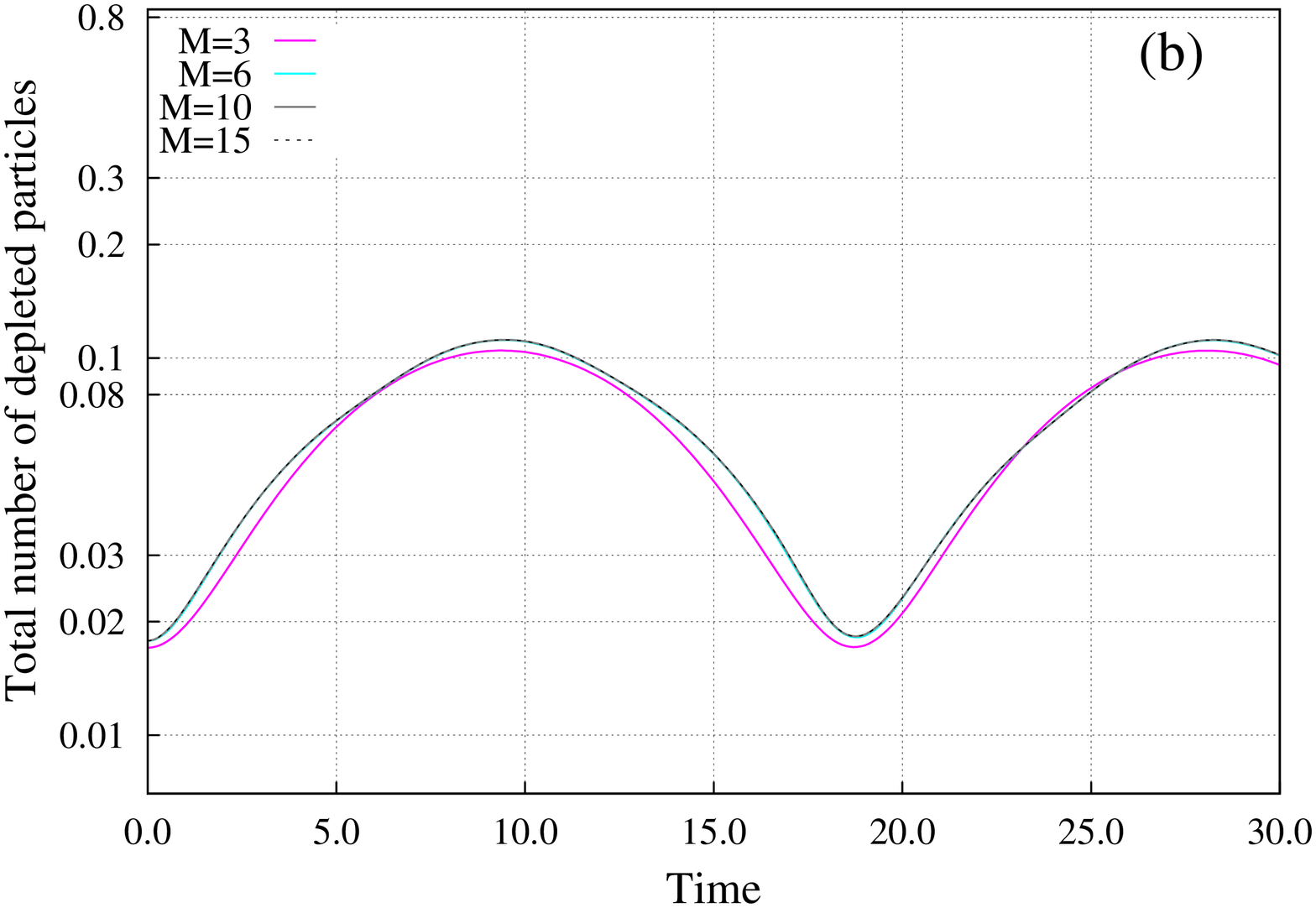}
\includegraphics[width=0.2930\columnwidth,angle=-90]{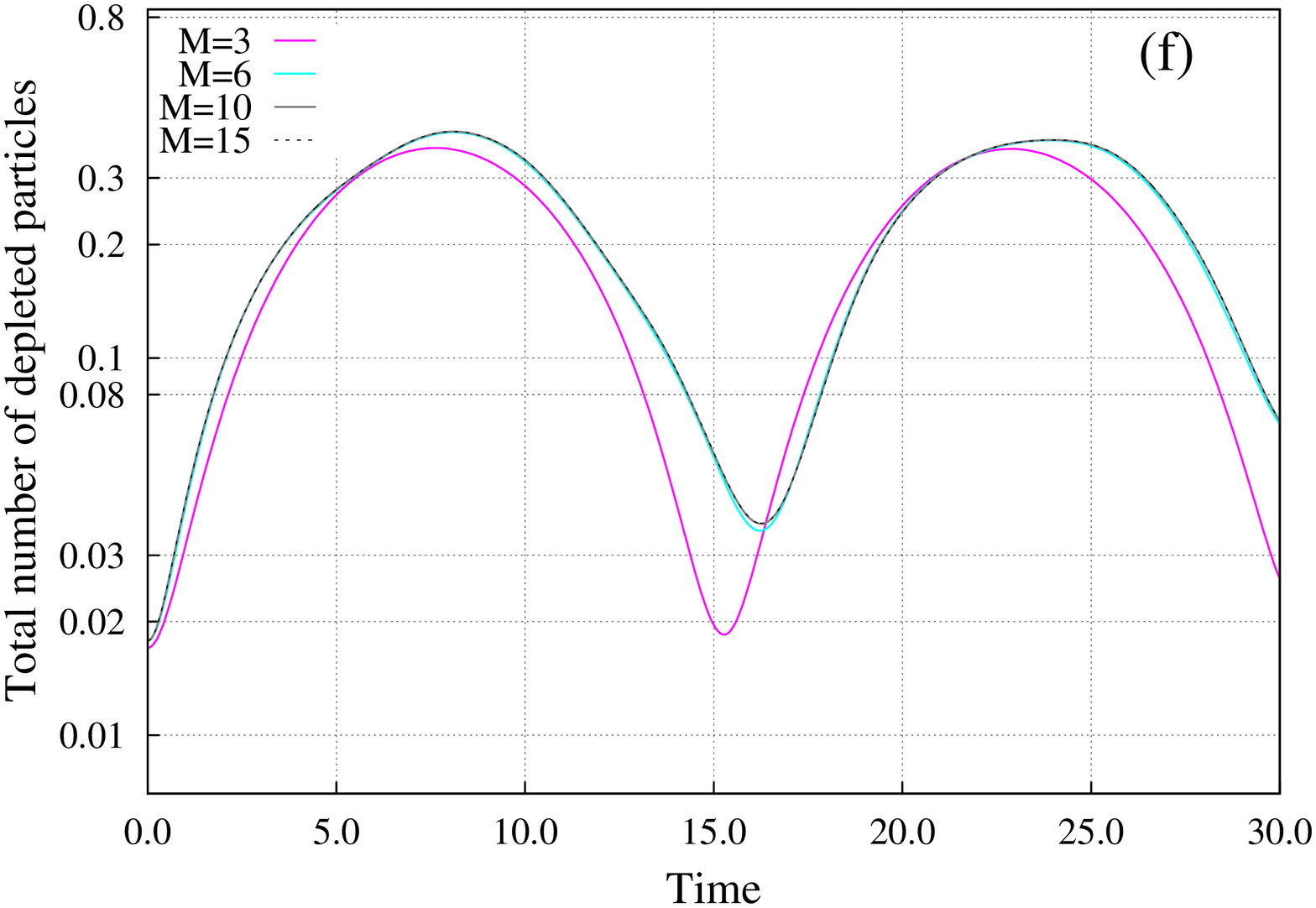}
\vglue 0.25 truecm
\hglue -1.0 truecm
\includegraphics[width=0.2930\columnwidth,angle=-90]{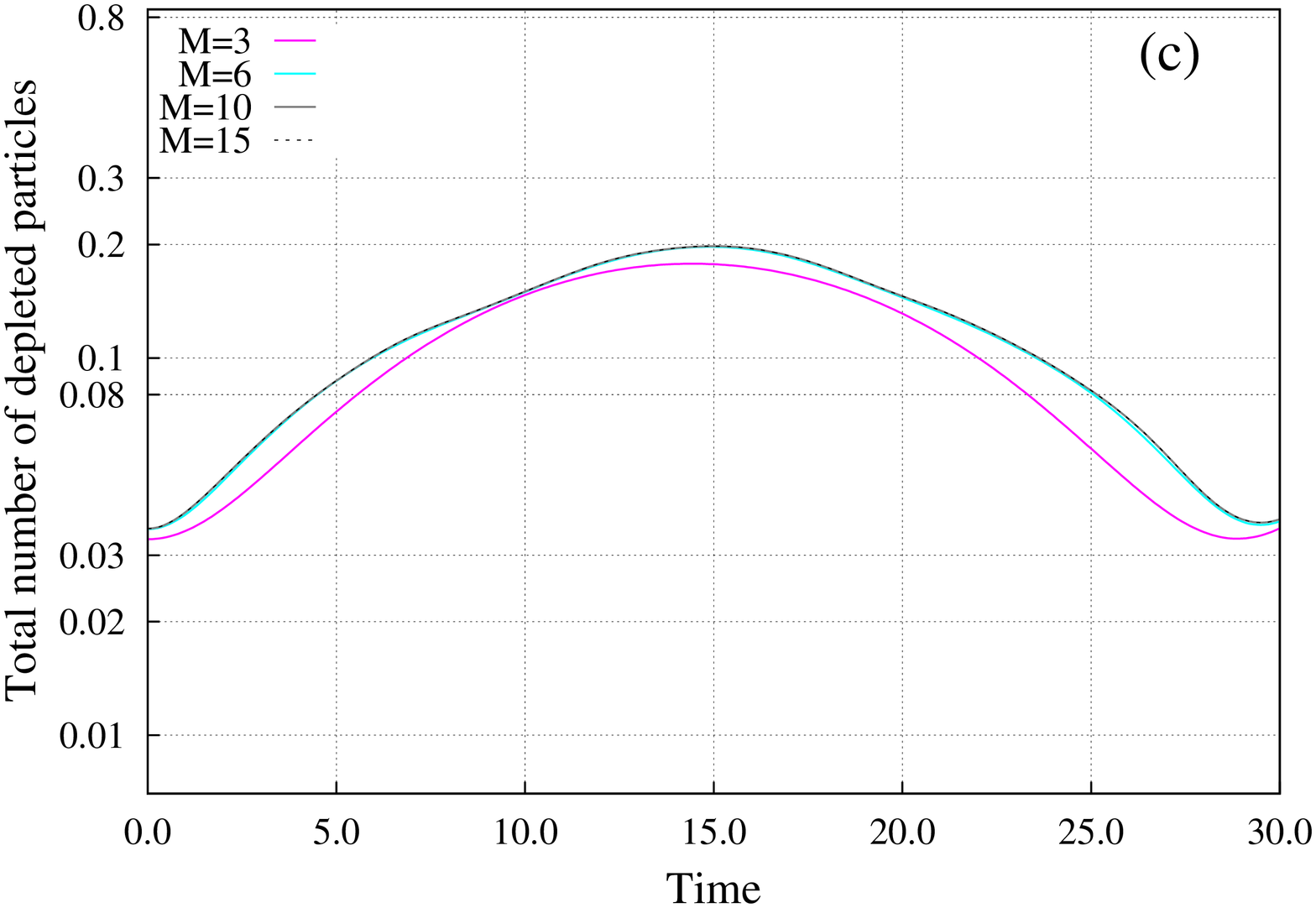}
\includegraphics[width=0.2930\columnwidth,angle=-90]{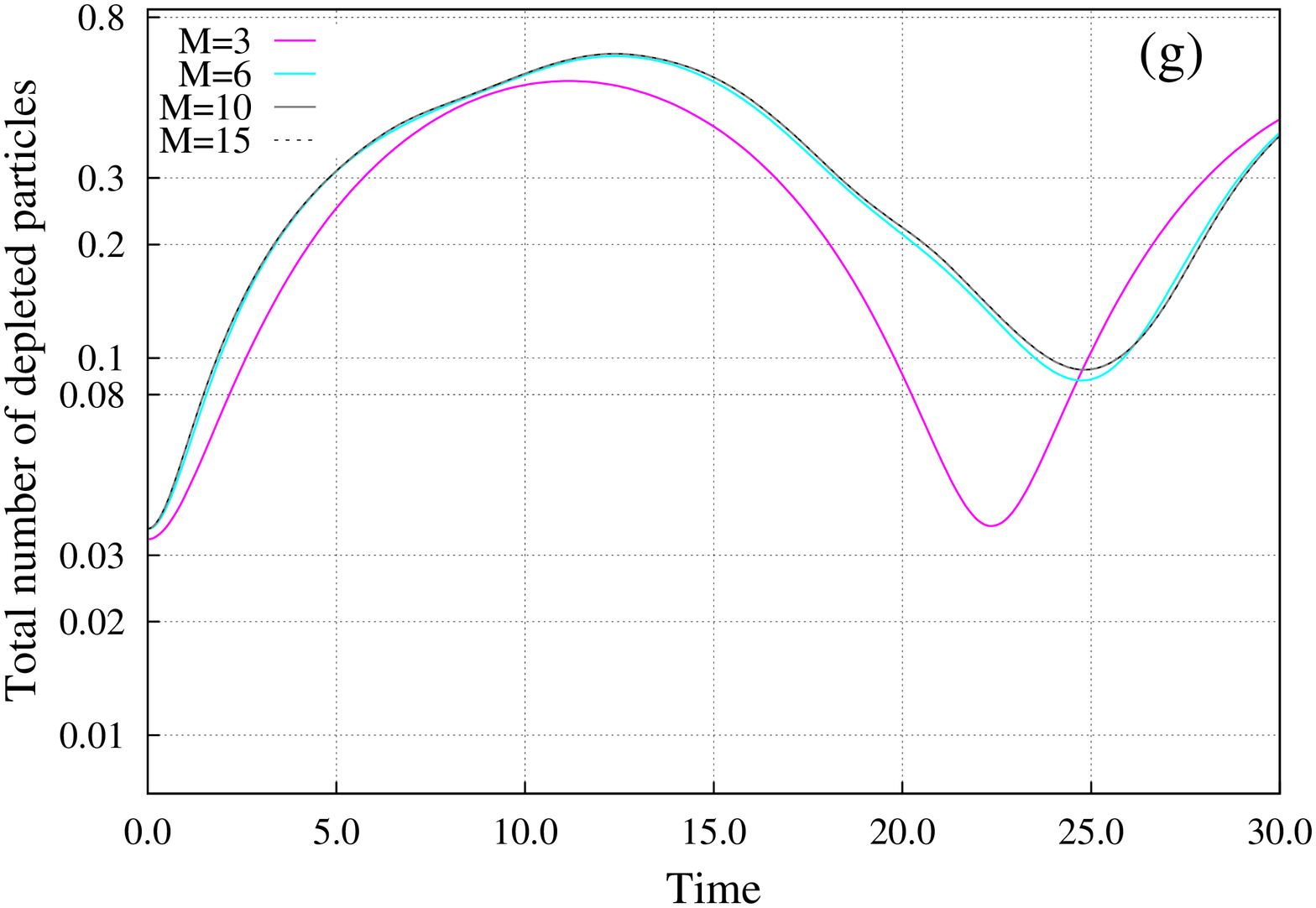}
\vglue 0.25 truecm
\hglue -1.0 truecm
\includegraphics[width=0.2930\columnwidth,angle=-90]{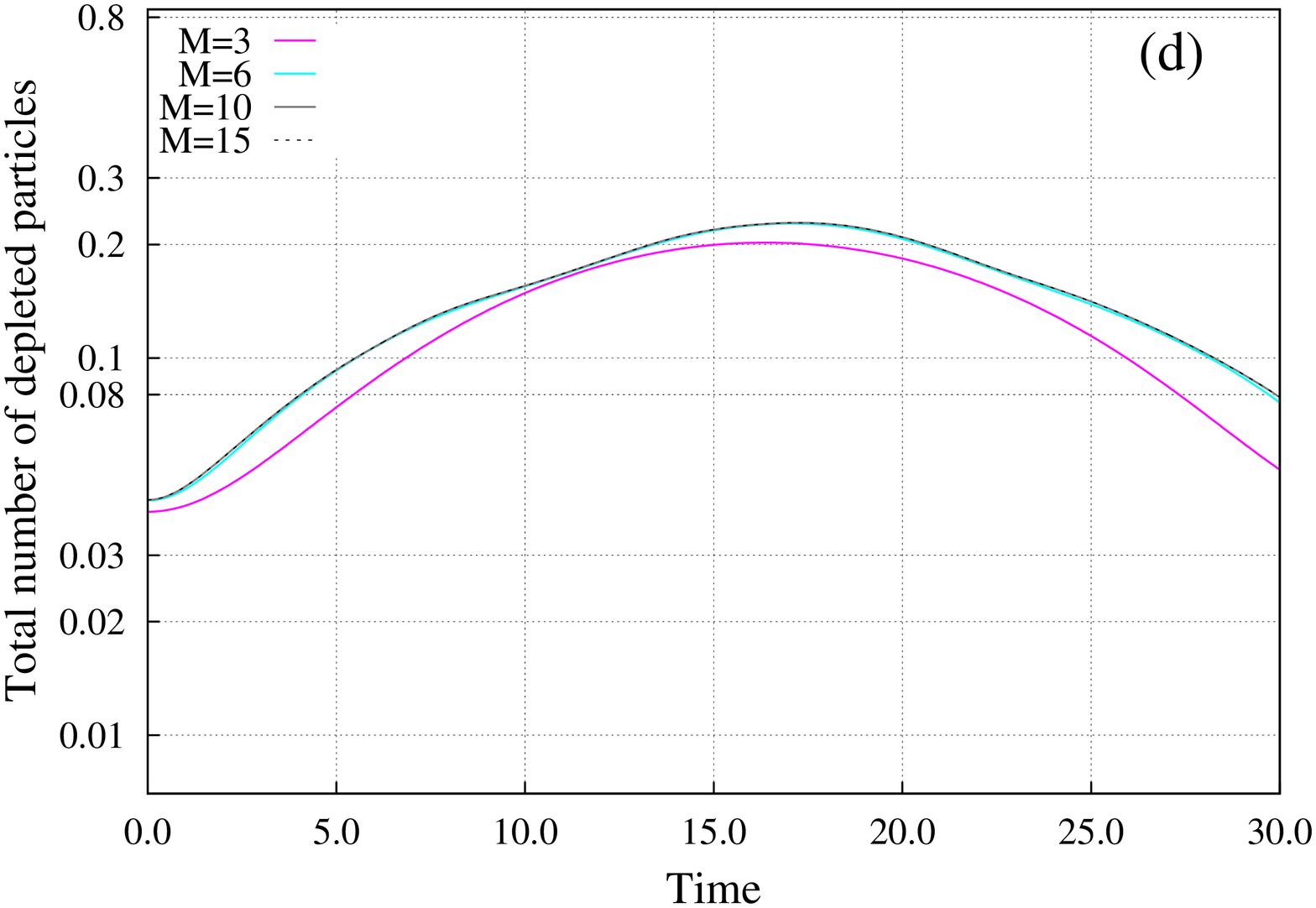}
\includegraphics[width=0.2930\columnwidth,angle=-90]{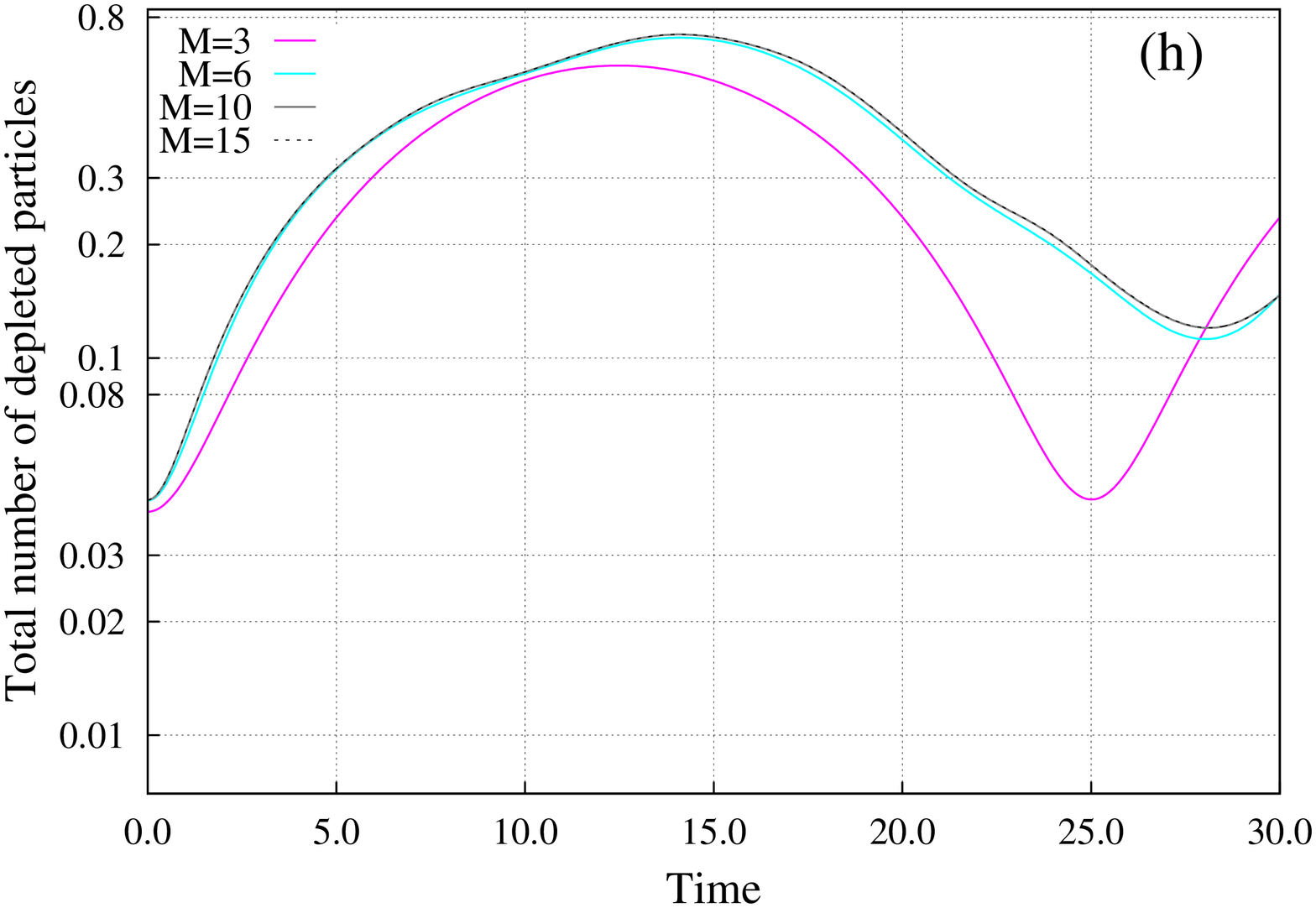}
\end{center}
\vglue -0.25 truecm
\caption{Depletion breathing dynamics following interaction quenches.
The total number of depleted particles, $N-n_1$, of $N=10$ bosons
following an interaction quench from $\lambda_0=0.02$ to $\lambda_0=0.04$
[left column, panels (a), (b), (c), and (d)]
and from $\lambda_0=0.02$ to $\lambda_0=0.08$
[right column, panels (e), (f), (g), and (h)].
The upper row [panels (a) and (e)] is for the annulus with barrier height $V_0=5$,
the second row [panels (b) and (f)] for $V_0=10$,
the third row [panels (c) and (g)] for $V_0=50$,
and the lower row [panels (d) and (h)] is for $V_0=100$.
$M=3$, $6$, $10$, and $15$ time-adaptive orbitals are used.
The respective variances are plotted for the two-fold interaction quench in 
Fig.~\ref{f3} and the four-fold quench in Fig.~\ref{f4}.
See the text for more details.
The quantities shown are dimensionless.}
\label{f5}
\end{figure}

Fig.~\ref{f3} records the time-dependent many-body variance per particle of $N=10$ bosons
in the different annuli for a two-fold interaction quench from $\lambda_0=0.02$ to $\lambda_0=0.04$
and, similarly, Fig.~\ref{f4} displays the results
for a four-fold interaction quench from $\lambda_0=0.02$ to $\lambda_0=0.08$.
Fig.~\ref{f5} collects the total number of depleted particles in all the above quench dynamics.
There are three major differences between the mean-field and many-body variances
which are the respective time-dependent values, the amplitudes of oscillations, and
their frequencies.

The energy pumped into the system as a result of the interaction quench is not much.
For the smallest $V_0=5$ annulus and the two-fold quench from $\lambda_0=0.02$ to $\lambda_0=0.04$ 
the energy per particle at the mean-field ($M=1$) and mean-body ($M=10$)
levels increases from $\frac{E}{N}=2.45107$ to $2.46863$ and from $\frac{E}{N}=2.45085$ to $2.46801$, respectively.
For the largest $V_0=100$ annulus and the four-fold quench from $\lambda_0=0.02$ to $\lambda_0=0.08$ 
the energy per particle at the $M=1$ and $M=10$
levels of theory increases, respectively, from $\frac{E}{N}=7.07186$ to $7.10565$
and from $\frac{E}{N}=7.07146$ to $7.10313$.

Let us examine the behavior of the variance, first at the mean-field level.
Upon quenching the interaction, the density performs breathing oscillations starting from its ground-state value
($\frac{1}{N} \Delta^2_{\hat X}=1.87$, $2.45$, $3.78$, and $4.37$ for $V_0=5$, $10$, $50$, and $100$, respectively).
The amplitude of oscillations is very small, see the blue curves in Figs.~\ref{f3} and \ref{f4}.
Furthermore, the amplitude of these small density oscillations further reduces when the barrier height is increased. 
This is since the annulus becomes narrower when its size increases.
For the largest annulus the density oscillations starting from $\frac{1}{N} \Delta^2_{\hat X}=4.37$
are hardly visible, see Figs.~\ref{f3}d and \ref{f4}d.
Corroborating the above is the frequency of the small-amplitude oscillations
which mildly increases with the barrier height, signifying that the annulus is becoming narrower with increasing size.
Finally, comparing the two-fold and four-fold quenches in Figs.~\ref{f3} and \ref{f4}, respectively,
we see that the latter hardly increases the small-amplitude oscillations and,
similarly, only slightly increases their frequency originating from the radial breathing excitation,
see further discussion below.

We now turn to
the many-particle position variance computed at the many-body level.
A global look at Figs.~\ref{f3} and \ref{f4} shows a qualitatively different behavior of the dynamics: 
The values are significantly smaller, and
the amplitudes of oscillations and their frequencies are substantially different.
There is a strong dependence on the barrier height $V_0$ 
and on whether the two-fold or the four-fold interaction quench is performed.
Explicitly, the time-dependent values of the variance can 
become less than $50\%$ from their initial values, see Fig.~\ref{f4}c and \ref{f4}d.
Accordingly, the amplitude of oscillations becomes larger than $50\%$.
These observation serve to identify and classify the many-body position variance as a two-dimensional quantity.
The frequency of oscillations shed further light on their origin.
Examining the dynamics in the four annuli sizes and both interaction quenches,
one learns that the frequency of oscillations strongly decreases with the annulus size, 
and mildly increases with the interaction-quench strength.
These suggest that angular excitations are involved,
that is a transfer of two bosons from the zero angular-momentum (highest-occupied) natural orbital $\phi_1(\r)=f(r)$
to a plus--and--minus unit angular-momentum (nearly unoccupied) degenerate natural orbitals $\phi_2(\r)=g(r)e^{+i\varphi}$
and $\phi_3(\r)=g(r)e^{-i\varphi}$ [$f(r)$ and $g(r)$ are radial functions which generally are not identical].
Indeed, such angular excitations in a ring-shaped trap decrease with the radius square
and increase with the interaction strength.
For instance, the frequency of oscillations increases by slightly more than twice
from the $V_0=5$ to $V_0=50$ annuli, see Figs.~\ref{f3}a,c and \ref{f4}a,c.
This is in accordance with the ratio of 
the above respective radii $R$ squares, $\frac{2.62(5)^2}{1.75(0)^2}=\frac{6.89}{3.06} = 2.25$.

Finally, to show that the angular excitations indeed contribute to the position variance,
we resort to its expression in terms of the natural orbitals,
$\frac{1}{N} \Delta^2_{\hat X} = \int d\r \frac{\rho(\r)}{N}x^2 - N\left[\int d\r \frac{\rho(\r)}{N}x\right]^2
+ \sum_{jpkq} \frac{\rho^{(2)}_{jpkq}}{N} \int d\r_1 d\r_2 \phi_j^\ast(\r_1) \phi_p^\ast(\r_2)  x_1x_2 \phi_k(\r_1) \phi_q(\r_2)$,
where 
$\rho^{(2)}(\r_1,\r_2,\r'_1,\r'_2) = \sum_{jpkq} \rho^{(2)}_{jpkq} \phi_j^\ast(\r'_1) \phi_p^\ast(\r'_2) \phi_k(\r_1) \phi_q(\r_2)$.
Transforming the last integral to polar coordinates,
one readily obtains a contribution from the above-described excitation channel,\break\hfill
$\frac{\rho^{(2)}_{3211}}{N} \int r_1 dr_1 d\varphi_1 r_2 dr_2 d\varphi_2
g^\ast(r_1) g^\ast(r_2) f(r_1) f(r_2) e^{-(\varphi_1-\varphi_2)} r_1 r_2 \frac{1}{2}\left[ \cos(\varphi_1-\varphi_2)+
\cos(\varphi_1+\varphi_2) \right]\break\hfill
=\frac{\rho^{(2)}_{3211}}{N} \left[\pi \int dr r^2 g^\ast(r) f(r) \right]^2$,
which is both symmetry allowed and, as Figs.~\ref{f3} and \ref{f4} depict, dominant.
We remind that these angular excitations are not available in the quench dynamics within mean-field theory,
but only radial excitations.
This is due to the restricted structure of the mean-field wavefunction
in comparison with the many-body wavefunction.

Finally, Fig.~\ref{f5} presents the time-dependent number of depleted particles 
of the $N=10$ bosons
in the annuli
for the two-fold $\lambda_0=0.02$ to $\lambda_0=0.04$
and four-fold $\lambda_0=0.02$ to $\lambda_0=0.08$ interaction quenches.
The frequencies of the number of depleted particles for the four annuli sizes and the two interactions quenches,
along with the amplitudes of their oscillations, nicely matches those of the many-body variances in Figs.~\ref{f3} and \ref{f4}.
Accordingly, this supports the above analysis and identification of angular excitations. 

\subsection{Larger systems}\label{LARGE_SYS}

\begin{figure}[!]
\begin{center}
\hglue -1.0 truecm
\includegraphics[width=0.345\columnwidth,angle=-90]{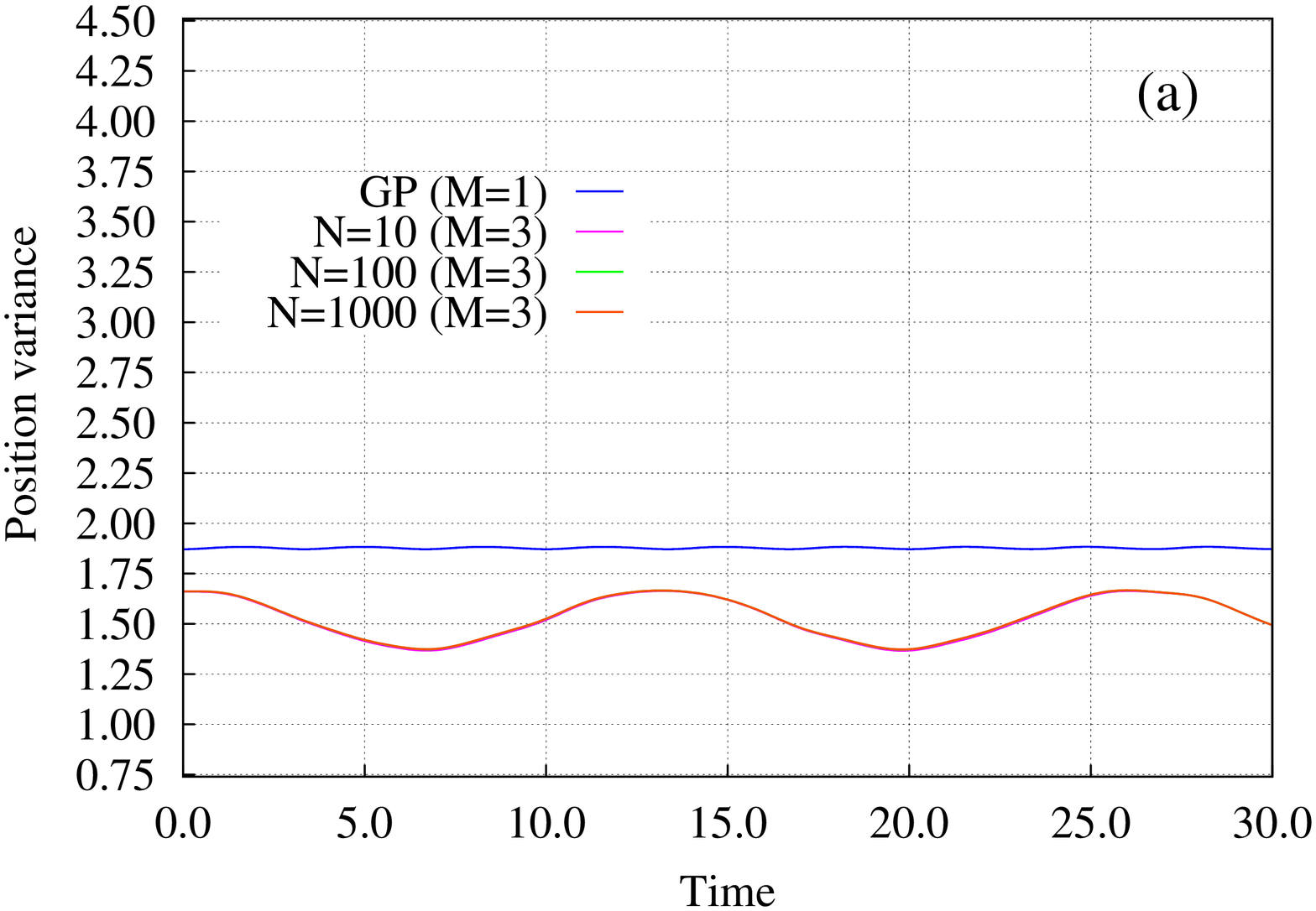}
\includegraphics[width=0.345\columnwidth,angle=-90]{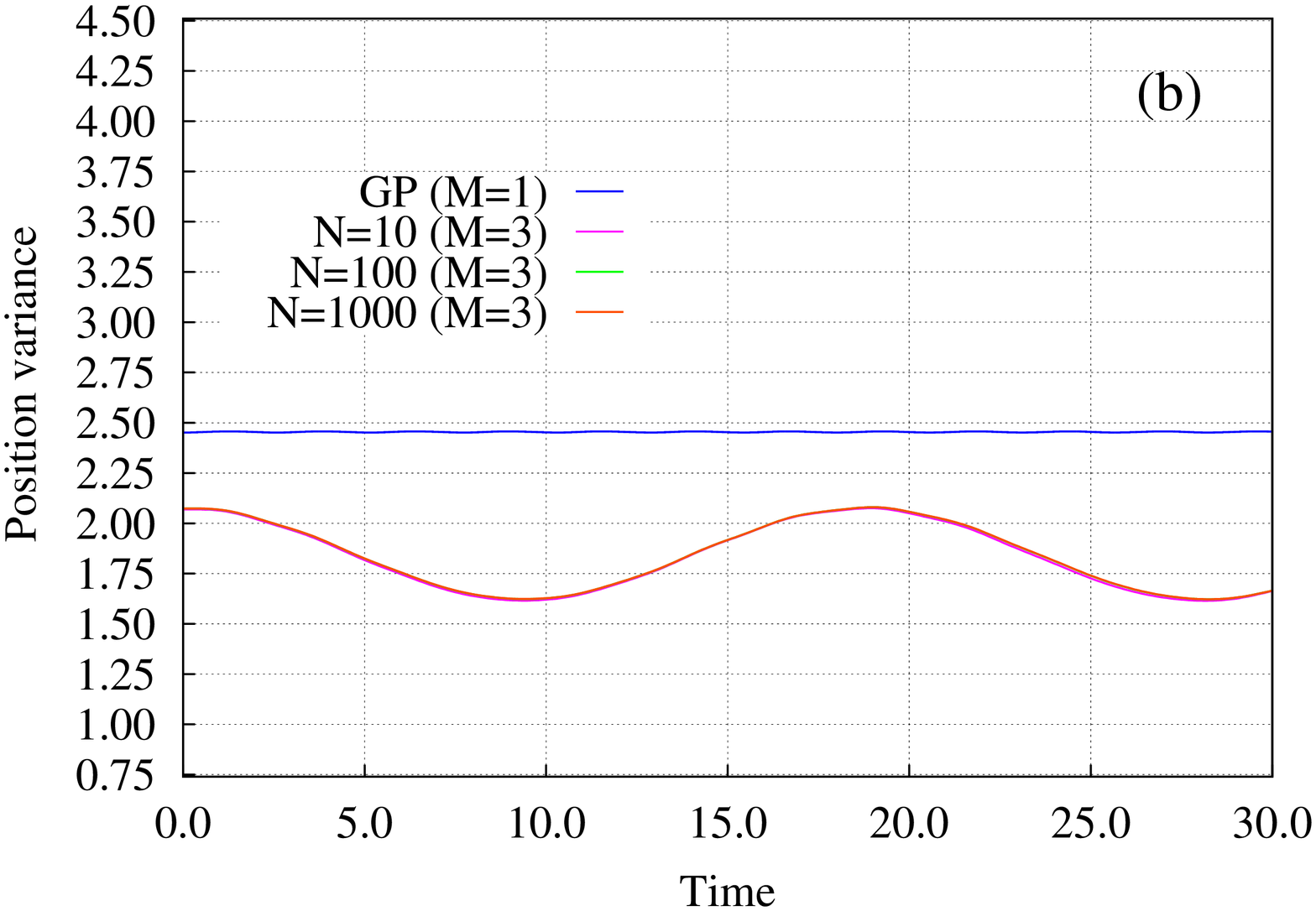}
\hglue -1.0 truecm
\includegraphics[width=0.345\columnwidth,angle=-90]{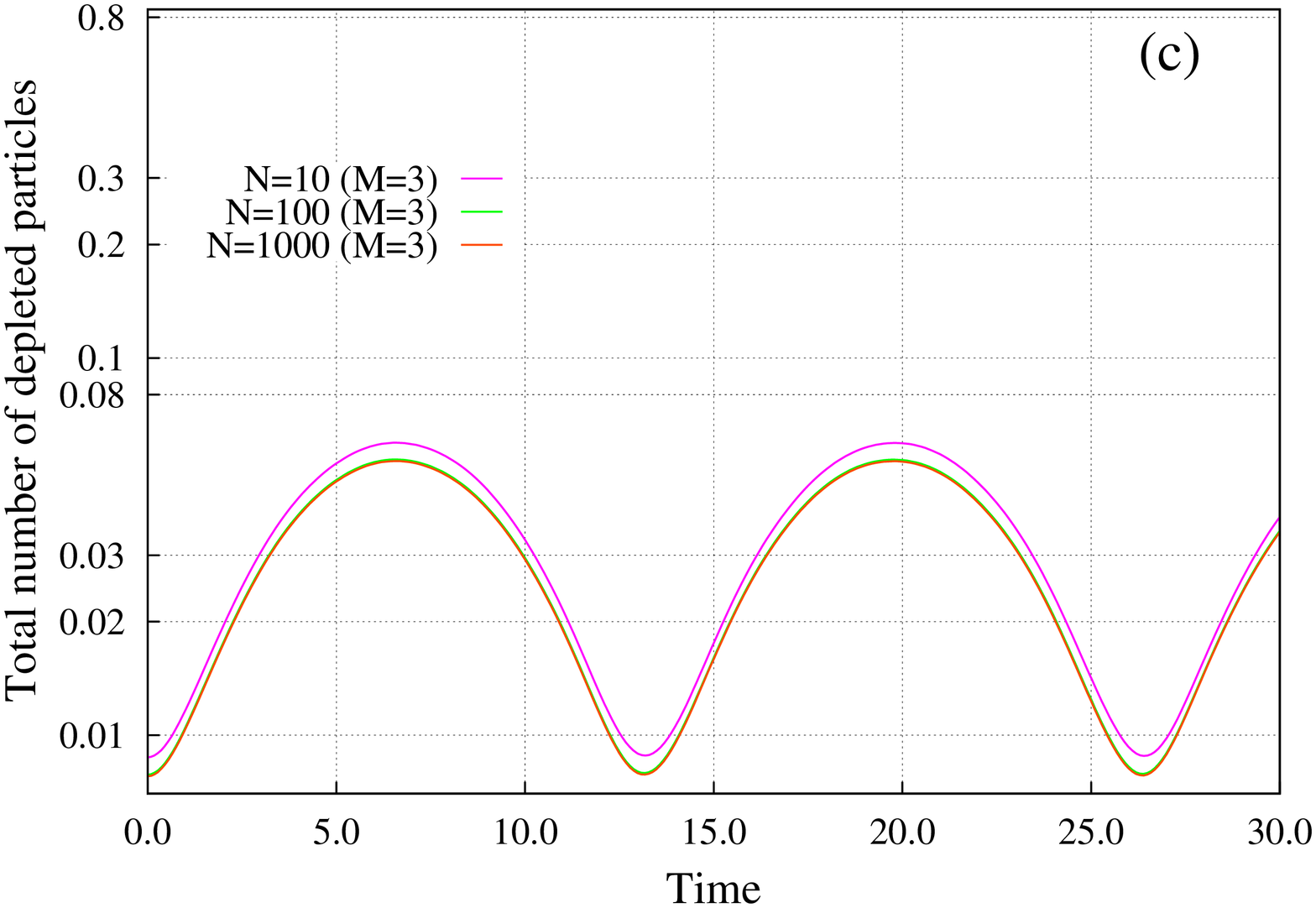}
\includegraphics[width=0.345\columnwidth,angle=-90]{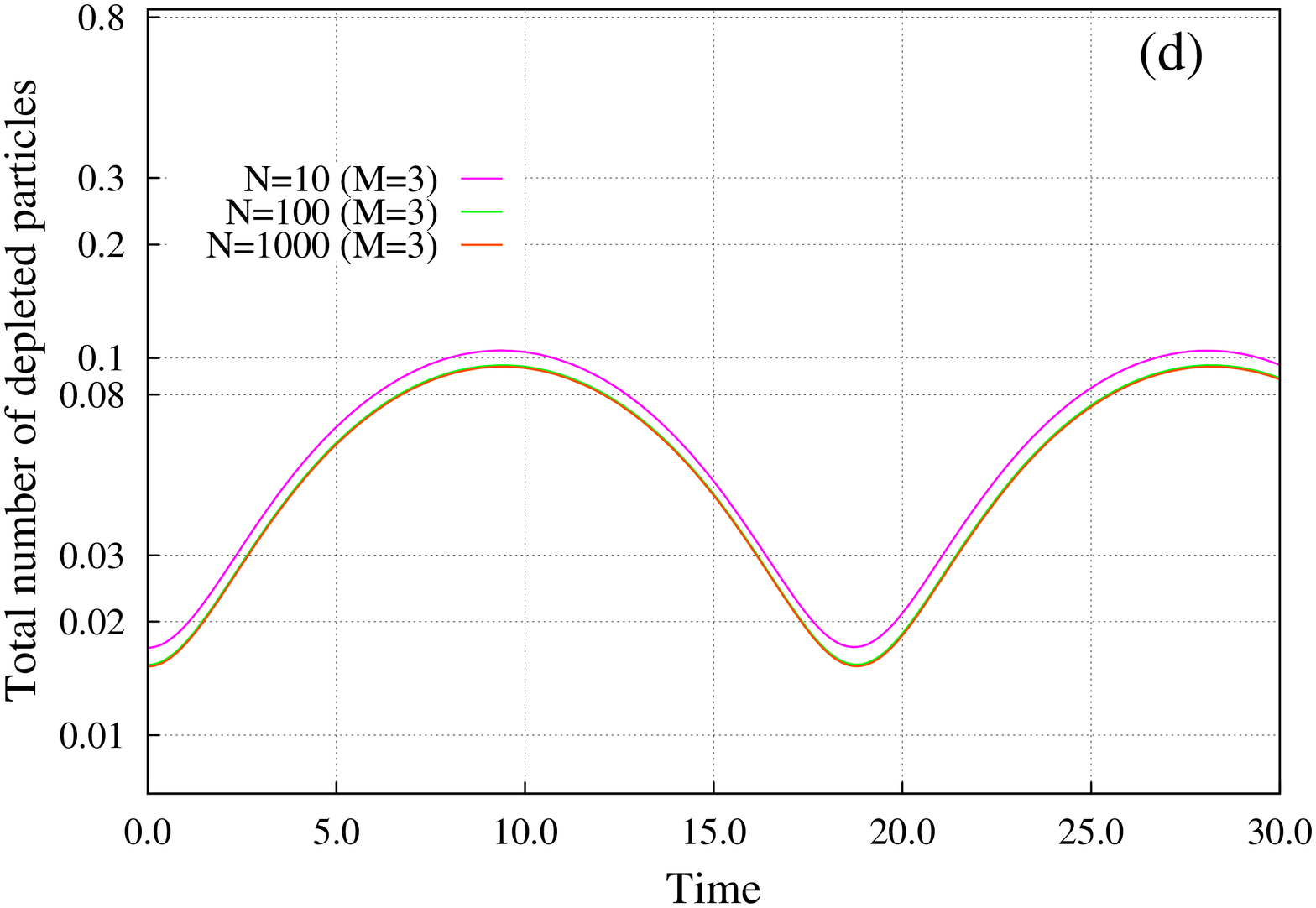}
\end{center}
\vglue 0.75 truecm
\caption{Breathing dynamics following a two-fold interaction quench {\it en route} to the infinite-particle limit.
The mean-field ($M=1$ time-adaptive orbitals) and many-body ($M=3$ time-adaptive orbitals)
position variances per particle,
$\frac{1}{N}\Delta^2_{\hat X}(t)$,
of $N=10$, $N=100$, and $N=1000$ bosons in the annuli with barrier heights (a) $V_0=5$ and (b) $V_0=10$
following a quench of the interaction parameter
from $\Lambda=\lambda_0(N-1)=0.18$ to $\Lambda=0.36$.
The respective total number of depleted particles $N-n_1$ is
plotted in panels (c) for $V_0=5$ and (d) for $V_0=10$.
See the text for more details.
The quantities shown are dimensionless.
}
\label{f6}
\end{figure}

So far we have dealt with the properties of a rather small number of bosons, $N=10$.
We see from the dynamics in Figs.~\ref{f3} and \ref{f4}
that, qualitatively, the many-body effect is obtained at the level of $M=3$ time-adaptive orbitals.
Yet, as the annulus size is increased along with the barrier height,
$M>3$ time-adaptive orbitals are needed to quantitatively and accurately describe the dynamics. 

For the two smaller rings, i.e., for barrier heights $V_0=5.0$ and $V_0=10.0$,
and the two-fold interaction quench from $\lambda_0=0.02$ to $\lambda_0=0.04$,
the above shows that the dynamics with $M=3$ time-adaptive orbitals
is already quantitative and accurate,
see Figs.~\ref{f3}a,b and Figs.~\ref{f5}a,b.
We therefore go to larger number of particles,
keeping the same interaction parameters for the quench, i.e.,
from $\Lambda=\lambda_0(N-1)=0.18$ to $\Lambda=0.36$.
Fig.~\ref{f6} collects the results for $N=10$, $N=100$, and $N=1000$ bosons.

We see that systems with the same interaction parameter $\Lambda$ and growing number of particles $N$
have essentially the same time-dependent position variance.
Importantly, the substantial differences between the many-body and mean-field variances persist, see Figs.~\ref{f6}a and \ref{f6}b.
Furthermore, the number of depleted particles is seen to converge to the same time-dependent behavior with increasing $N$,
see Figs.~\ref{f6}c and \ref{f6}d.
This implies the systems are becoming more and more condensed,
and eventually $100\%$ condensed at the infinite-particle limit.
The results provide strong evidence that for
ever growing numbers of particles and at a fixed interaction parameter
the dimensionality of the position variance and that of the density (mean-field variance)
can behave in a different manner,
the former being a two-dimensional quantity and
the latter an essentially quasi-one-dimensional one.

\section{Concluding Remarks}\label{CONCLUSIONS}

In conclusion, we have studied the ground state and out-of-equilibrium dynamics of BECs in two-dimensional annular traps
by solving the many-boson Schr\"odinger equation numerically accurately
using the MCTDHB method.
We focused in the present work on weakly-interacting BECs exhibiting small numbers of depleted particles.
Examining the mean-field position variance, which accounts for the shape of the radially-symmetric
density only,
and the many-body position variance, which incorporates tiny angular excitations through the reduced two-particle density matrix,
shows that, 
whereas the density can behave in a quasi-one-dimensional manner,
the variance in contrast can `live' in the two-dimensional plane.
The found dimensionality effect persists for larger number of bosons and the same interaction parameter
{\it en route} to the infinite-particle limit,
where the bosons become $100\%$ condensed.
This is appealing, and adds another dimension
to the growing dissonance between
the variance and density of a trapped BEC.

\section*{Acknowledgements}

This paper is dedicated to Professor Nimrod Moiseyev, $\mathrm{\overset{>}{N}}$imrod,
a dear friend for long time and my advisor,
on the occasion of his 70th birthday.
This research was supported by the Israel Science Foundation (Grant No. 600/15).
We thank Raphael Beinke, Sudip Haldar, Lorenz Cederbaum, Kaspar Sakmann, Shachar Klaiman, and Alexej Streltsov for discussions.
Computation time on the BwForCluster and the Cray XC40 
system Hazelhen at the High Performance Computing Center
Stuttgart (HLRS) is gratefully acknowledged.

\appendix
\section{Many-body momentum variance}\label{MOMENTUM}

\begin{figure}[!]
\begin{center}
\hglue -1.25 truecm
\includegraphics[width=0.2630\columnwidth,angle=-90]{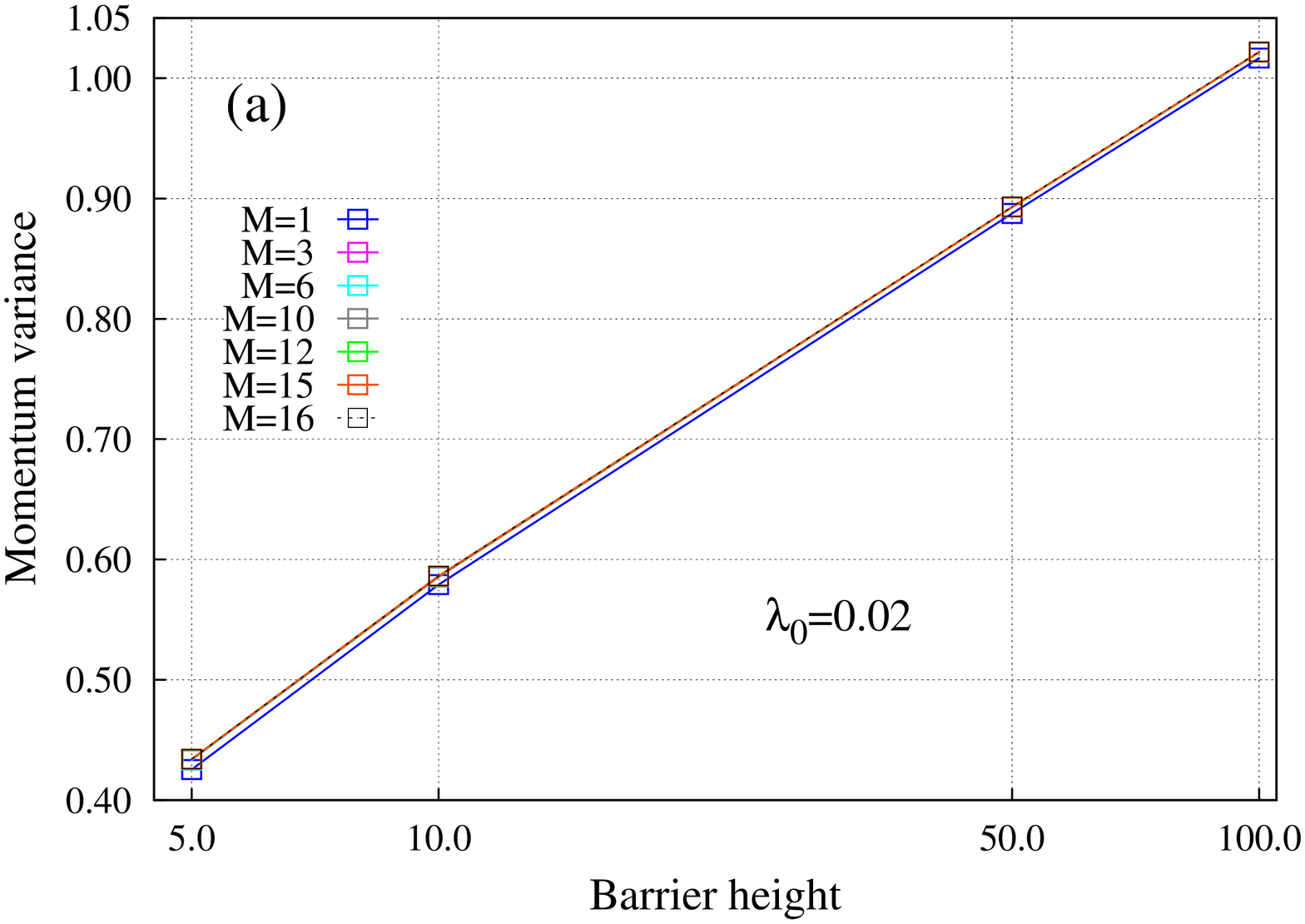}\hglue -0.25 truecm
\includegraphics[width=0.2630\columnwidth,angle=-90]{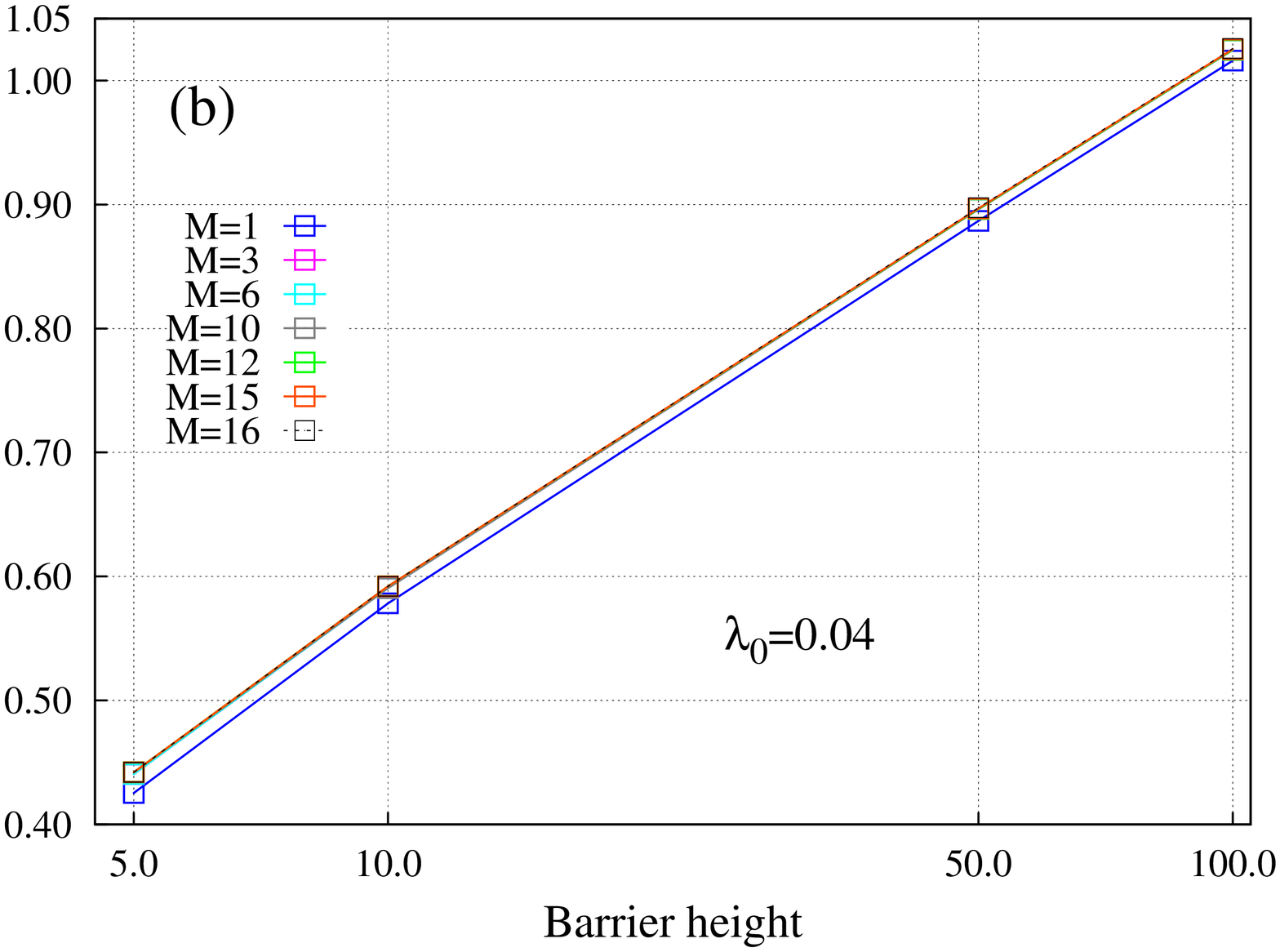}\hglue -0.25 truecm
\includegraphics[width=0.2630\columnwidth,angle=-90]{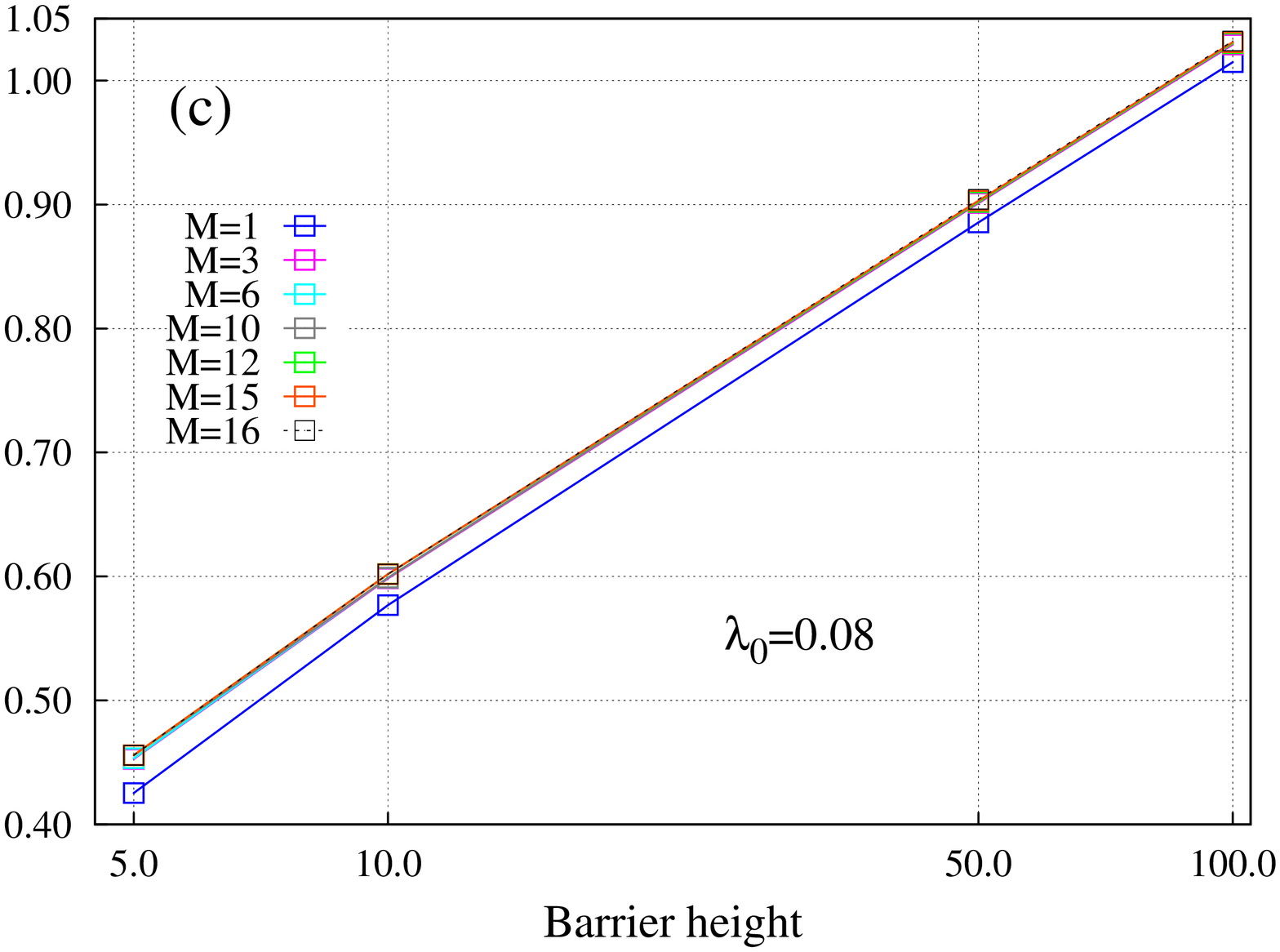}
\end{center}
\vglue 0.75 truecm
\caption{Ground-state properties as a function of the barrier height $V_0$ and interaction strength $\lambda_0$: 
Many-particle momentum variance per particle, $\frac{1}{N}\Delta^2_{\hat P_X}$.
The number of bosons is $N=10$.
The barrier heights are $V_0=5$, $10$, $50$, and $100$.
The interaction strengths are (a) $\lambda_0=0.02$, (b) $\lambda_0=0.04$,
and (c) and $\lambda_0=0.08$.
Actual data are marked by symbols, the continuous curves are to guide the eye only.
See the text for more details.
The quantities shown are dimensionless.}
\label{f7}
\end{figure}

\begin{figure}[!]
\begin{center}
\hglue -1.65 truecm
\includegraphics[width=0.385\columnwidth,angle=-90]{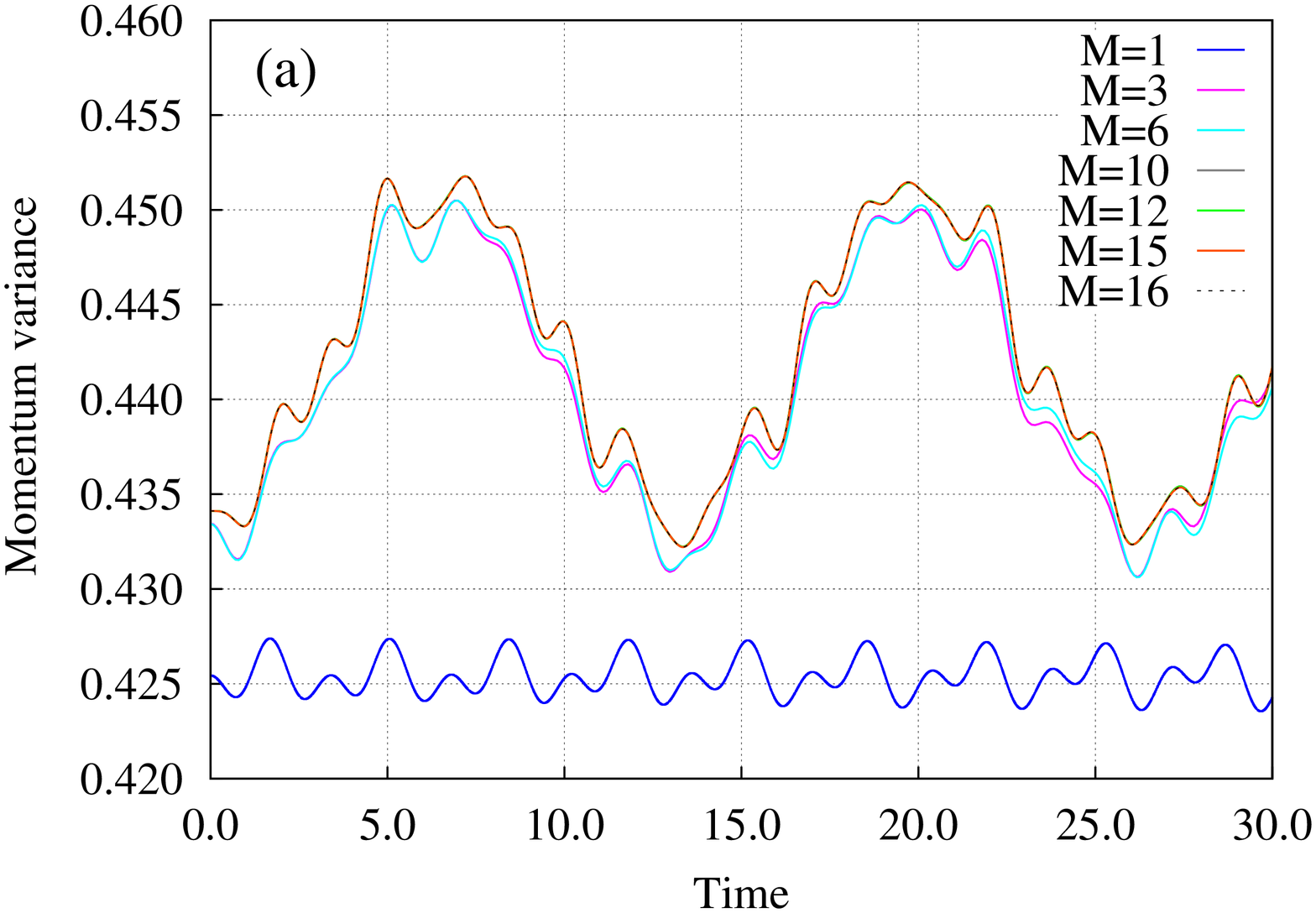}
\hglue -0.15 truecm
\includegraphics[width=0.385\columnwidth,angle=-90]{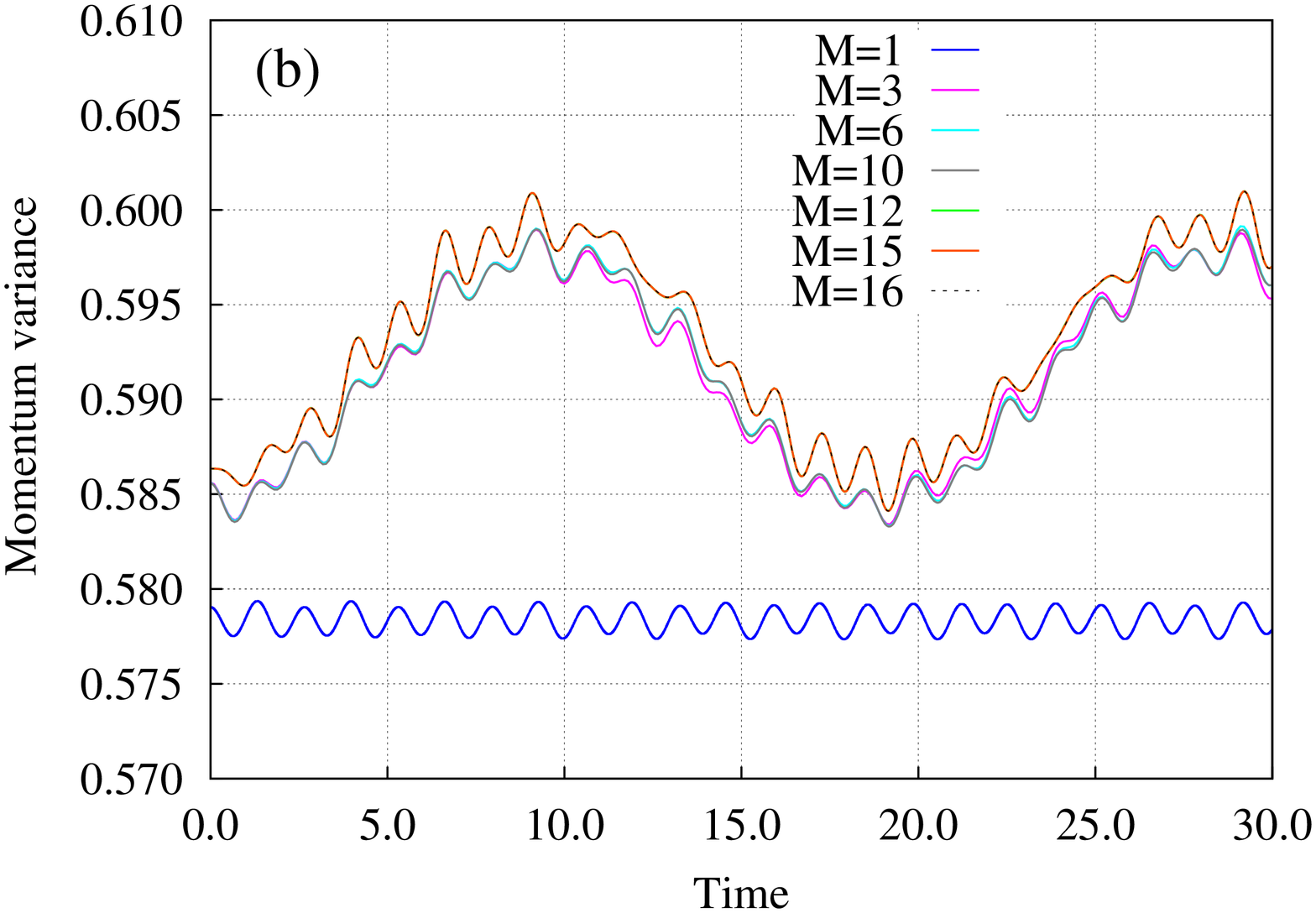}
\end{center}
\vglue 0.75 truecm
\caption{Variance breathing dynamics following a two-fold interaction quench:
Many-particle momentum variance per particle, $\frac{1}{N}\Delta^2_{\hat P_X}(t)$.
The mean-field ($M=1$ time-adaptive orbitals) and many-body ($M=3$, $6$, $10$, $12$, $15$, and $16$ time-adaptive orbitals)
results for $N=10$ bosons in the annuli with
barrier heights (a) $V_0=5$ and (b) $V_0=10$
following an interaction quench from $\lambda_0=0.02$ to $\lambda_0=0.04$.
See the text for more details.
The quantities shown are dimensionless.}
\label{f8}
\end{figure}

We collect complementary results of the momentum variance per particle,
$\frac{1}{N}\Delta^2_{\hat P_X}(t)$ with $\hat P_X = \sum_{j=1}^N \frac{1}{i}\frac{\partial}{\partial x_j}$,
in this appendix.
Fig.~\ref{f7} shows the statics
and Fig.~\ref{f8} the dynamics.

Unlike the position variance, which exhibits large differences between its many-body and mean-field descriptions,
the respective differences for the momentum variance are much smaller.
In Fig.~\ref{f7} we depict the momentum variance of the ground state of $N=10$ bosons 
in the above four annuli with barrier heights $V_0=5$, $10$, $50$, and $100$
for the three interaction strengths $\lambda_0=0.02$, $0.04$, and $0.08$.
The curves for the mean-field and many-body momentum variance of the ground state
almost sit atop each other for $\lambda_0=0.02$, see Fig.~\ref{f7}a, and seen to slightly part away from each other
for the stronger interactions $\lambda_0=0.04$ and $\lambda_0=0.08$, see Figs.~\ref{f7}b and \ref{f7}c.
Explicitly, 
for the smallest $V_0=5$ annulus and weakest interaction $\lambda_0=0.02$,
$\frac{1}{N}\Delta^2_{\hat P_X}=0.42$ ($M=1$) and $\frac{1}{N}\Delta^2_{\hat P_X}=0.43$ ($M=15$),
and for the largest $V_0=100$ annulus and strongest interaction $\lambda_0=0.08$,
$\frac{1}{N}\Delta^2_{\hat P_X}=1.01$ ($M=1$) and $\frac{1}{N}\Delta^2_{\hat P_X}=1.03$ ($M=15$).
Further, it is seen that the momentum variance increases with the annulus size, see Fig.~\ref{f7}.

In Fig.~\ref{f8} we depict the time-dependent momentum variance of $N=10$ bosons
for the two-fold interaction quench from $\lambda_0=0.02$ to $\lambda_0=0.04$
in the two smaller rings, i.e., for barrier heights $V_0=5.0$ and $V_0=10.0$.
Globally, and similarly to static results, Fig.~\ref{f7}, the differences in 
absolute values between the many-body and mean-field results are only a couple of percents.
Again, this is in contrast to the corresponding
time-dependent position variances which exhibit tens of percents differences from each other.
However, examining the fine-structure differences between the many-body and mean-field quantities indicates their origin.
At the mean-field level, $\frac{1}{N}\Delta^2_{\hat P_X}(t)$ performs small-amplitude, high-frequency oscillations.
Their frequency is slightly increases and amplitude mildly decreases with the annulus size,
compare Figs.~\ref{f8}a and \ref{f8}b.
These mean-field oscillations are reminiscent of the radial breathing oscillations of the density.
At the many-body level, $\frac{1}{N}\Delta^2_{\hat P_X}(t)$ also exhibits such small-amplitude, high-frequency oscillations,
but they are dressed by relatively larger-amplitude, lower-frequency oscillations.
The frequency of the latter grows with the annulus size and their amplitude mildly decreases.
These many-body oscillations reflect the angular excitations discussed above,
compare to Figs.~\ref{f3}a,b and Figs.~\ref{f5}a,b.
 
Recapitulating,
the many-particle momentum variance, $\frac{1}{N}\Delta^2_{\hat P_X}$, 
of BECs in annuli does not exhibit in momentum space analogous
dimensionality effects to the many-particle position variance, $\frac{1}{N}\Delta^2_{\hat X}$, in real space,
when compared at the many-body and mean-field levels.
It would be interesting to search for situations when $\frac{1}{N}\Delta^2_{\hat P_X}$ does exhibit such effects.  



\begin{thebibliography}{99}

\bibitem{rev1} E. A. Cornell and C. E. Wieman,
                     {\rm Nobel Lecture: Bose-Einstein condensation in a dilute gas, 
                     the first 70 years and some recent experiments},
                     Rev. Mod. Phys. {\bf 74}, 875 (2002).

\bibitem{rev2} W. Ketterle, 
                     {\rm Nobel lecture: When atoms behave as waves:
                     Bose-Einstein condensation and the atom laser},
                     Rev. Mod. Phys. {\bf 74}, 1131 (2002).

\bibitem{rev3} F. Dalfovo, S. Giorgini, L. P. Pitaevskii, and S. Stringari,
                     {\rm Theory of Bose-Einstein condensation in trapped gases},
                     Rev. Mod. Phys. {\bf 71}, 463 (1999). 

\bibitem{rev4} A. J. Leggett,
                     {\rm Bose-Einstein condensation in the alkali gases: Some fundamental concepts},
                     Rev. Mod. Phys. {\bf 73}, 307 (2001).  

\bibitem{rev5} I. Bloch, J. Dalibard, and W. Zwerger,
                     {\rm Many-body physics with ultracold gases},
                     Rev. Mod. Phys. {\bf 80}, 885 (2008).

\bibitem{INF1} Y. Castin and R. Dum, 
                          {\rm Low-temperature Bose-Einstein condensates in time-dependent traps: 
                          Beyond the U(1) symmetry breaking approach}, 
                          Phys. Rev. A {\bf 57}, 3008 (1998).

\bibitem{INF2} E. H. Lieb, R. Seiringer, and J. Yngvason,
                      {\rm Bosons in a trap: A rigorous derivation of the
                      Gross-Pitaevskii energy functional},
                      Phys. Rev. A {\bf 61}, 043602 (2000).

\bibitem{INF3} E. H. Lieb and R. Seiringer,
                       {\rm Proof of Bose-Einstein Condensation for Dilute Trapped Gases},
                       Phys. Rev. Lett. {\bf 88}, 170409 (2002).

\bibitem{INF4} L. Erd\H{o}s, B. Schlein, and H.-T. Yau,
                       {\rm Rigorous Derivation of the Gross-Pitaevskii Equation}, 
                       Phys. Rev. Lett. {\bf 98}, 040404 (2007).

\bibitem{INF5} L. Erd\H{o}s, B. Schlein, and H.-T. Yau,
                      {\rm Derivation of the cubic non-linear Schr\"odinger equation 
                      from quantum dynamics of many-body systems},
                      Invent. Math. {\bf 167}, 515 (2007).

\bibitem{INF6} L. S. Cederbaum,
                        {\rm Exact many-body wave function and properties
                        of trapped bosons in the infinite-particle limit},
                        Phys. Rev. A {\bf 96}, 013615 (2017).

\bibitem{var1} S. Klaiman and O. E. Alon,
                      {\rm Variance as a sensitive probe of correlations}, 
                      Phys. Rev. A {\bf 91}, 063613 (2015).

\bibitem{var2} S. Klaiman, A. I. Streltsov, and O. E. Alon, 
                      {\rm Uncertainty product of an out-of-equilibrium many-particle system},
                      Phys. Rev. A {\bf 93}, 023605 (2016).

\bibitem{overlap} S. Klaiman and L. S. Cederbaum,
                          {\rm Overlap of exact and Gross-Pitaevskii wave functions in
                          Bose-Einstein condensates of dilute gases},
                          Phys. Rev. A {\bf 94}, 063648 (2016).

\bibitem{var3} K. Sakmann and J. Schmiedmayer,
                     {\rm Conserving symmetries in Bose-Einstein condensate dynamics requires many-body theory},
                     arXiv:1802.03746v2 [cond-mat.quant-gas].

\bibitem{var4} S. Klaiman, R. Beinke, L. S. Cederbaum, A. I. Streltsov, and O. E. Alon,
                      {\rm Variance of an anisotropic Bose-Einstein condensate},
                      Chem. Phys. {\bf 509}, 45 (2018).

\bibitem{var_ap1} M. Theisen and A. I. Streltsov,
                          {\rm Many-body excitations and deexcitations in trapped ultracold bosonic clouds},
                          Phys. Rev. A {\bf 94}, 053622 (2016).

\bibitem{rapha_excite} R. Beinke, L. S. Cederbaum, O. E. Alon,
                                 {\rm  Enhanced many-body effects in the excitation spectrum
                                 of a weakly interacting rotating Bose-Einstein condensate},
                                 Phys. Rev. A {\bf 98}, 053634 (2018).

\bibitem{var_ap2} S. K. Haldar and O. E. Alon,
                         {\rm Impact of the range of the interaction on the quantum
                         dynamics of a bosonic Josephson junction},
                         Chem. Phys. {\bf 509}, 72 (2018).

\bibitem{brand_1} J. G. Cosme, C. Weiss, and J. Brand,
                           {\rm Center-of-mass motion as a sensitive convergence test for 
                           variational multimode quantum dynamics},
                           Phys. Rev. A {\bf 94}, 043603 (2016).

\bibitem{attractive_2} O. E. Alon and L. S. Cederbaum,
                                {\rm Attractive Bose-Einstein condensates in anharmonic traps:
                                Accurate numerical treatment and the intriguing physics of the variance},
                                Chem. Phys. {\bf 515}, 287 (2018).

\bibitem{rn1} K. Sakmann, A. I. Streltsov, O. E. Alon, and L. S. Cederbaum,
                    {\rm Exact ground state of finite Bose-Einstein condensates on a ring},
                    Phys. Rev. A {\bf 72}, 033613 (2005).

\bibitem{rn2} S. Gupta, K. W. Murch, K. L. Moore, T. P. Purdy, and D. M. Stamper-Kurn,
                    {\rm Bose-Einstein Condensation in a Circular Waveguide},
                    Phys. Rev. Lett. {\bf 95}, 143201 (2005).
                  
\bibitem{rn3} M. Cozzini, B. Jackson, and S. Stringari,
                    {\rm Vortex signatures in annular Bose-Einstein condensates},
                    Phys. Rev. A {\bf 73}, 013603 (2006).

\bibitem{rn4} C. G. Bao,
                     {\rm Oscillation bands of Bose-Einstein condensates on a ring: Beyond the mean-field theory},
                      Phys. Rev. A {\bf 75}, 063626 (2007).

\bibitem{rn5} J. Smyrnakis, S. Bargi, G. M. Kavoulakis, M. Magiropoulos, K. K\"arkk\"ainen, and S. M. Reimann,
                    {\rm Mixtures of Bose Gases Confined in a Ring Potential},
                    Phys. Rev. Lett. {\bf 103}, 100404 (2009).

\bibitem{rn6} P. L. Halkyard, M. P. A. Jones, and S. A. Gardiner,
                    {\rm Rotational response of two-component Bose-Einstein condensates in ring traps},
                    Phys. Rev. A {\bf 81}, 061602(R) (2010).

\bibitem{rn7} L. Mathey, A. Ramanathan, K. C. Wright, S. R. Muniz, W. D. Phillips, and C. W. Clark,
                    {\rm Phase fluctuations in anisotropic Bose-Einstein condensates: From cigars to rings},
                    Phys. Rev. A {\bf 82}, 033607 (2010).

\bibitem{rn8} B. E. Sherlock, M. Gildemeister, E. Owen, E. Nugent, and C. J. Foot,
                    {\rm Time-averaged adiabatic ring potential for ultracold atoms},
                     Phys. Rev. A {\bf 83}, 043408 (2011).

\bibitem{rn9} S. Z\"ollner, G. M. Bruun, C. J. Pethick, and S. M. Reimann,
                      {\rm Bosonic and Fermionic Dipoles on a Ring},
                      Phys. Rev. Lett. {\bf 107}, 035301 (2011).

\bibitem{rn10} R. Dubessy, T. Liennard, P. Pedri, and H. Perrin,
                      {\rm Critical rotation of an annular superfluid Bose-Einstein condensate},                      
                      Phys. Rev. A {\bf 86}, 011602(R) (2012).

\bibitem{rn11} S. J. Woo and Y.-W. Son,
                      {\rm Vortex dynamics in an annular Bose-Einstein condensate},
                      Phys. Rev. A {\bf 86}, 011604(R) (2012).

\bibitem{rn12} S. Moulder, S. Beattie, R. P. Smith, N. Tammuz, and Z. Hadzibabic,
                      {\rm Quantized supercurrent decay in an annular Bose-Einstein condensate},
                      Phys. Rev. A {\bf 86}, 013629 (2012).

\bibitem{rn13} L. A. Toikka and K.-A. Suominen,
                      {\rm Snake instability of ring dark solitons in toroidally trapped Bose-Einstein condensates},
                      Phys. Rev. A {\bf 87}, 043601 (2013).

\bibitem{rn14} S. Eckel, J. G. Lee, F. Jendrzejewski, N. Murray, C. W. Clark, C. J. Lobb, W. D. Phillips, M. Edwards, and G. K. Campbell,
                    {\rm Hysteresis in a quantized superfluid ‘atomtronic’ circuit},
                    Nature (London) {\bf 506}, 200 (2014).

\bibitem{rn15} A. Mu\~noz Mateo, A. Gallem\'i, M. Guilleumas, and R. Mayol,
                      {\rm Persistent currents supported by solitary waves in toroidal Bose-Einstein condensates},
                      Phys. Rev. A {\bf 91}, 063625 (2015).

\bibitem{rn16} M. Kol\'a\v{r}, T. Opatrn\'y, and K. K. Das,
                     {\rm Criticality and spin squeezing in the rotational dynamics of a Bose-Einstein condensate on a ring lattice},
                     Phys. Rev. A {\bf 92}, 043630 (2015).

\bibitem{rn17} A. Roy and D. Angom,
                     {\rm Geometry-induced modification of fluctuation spectrum in quasi-two-dimensional condensates},
                     New J. Phys. {\bf 18}, 083007 (2016).

\bibitem{rn18} A. Roussou, J. Smyrnakis, M. Magiropoulos, N. K. Efremidis, and G. M. Kavoulakis,
                      {\rm Rotating Bose-Einstein condensates with a finite number of atoms confined in a ring potential:
                      Spontaneous symmetry breaking beyond the mean-field approximation},
                      Phys. Rev. A {\bf 95}, 033606 (2017).

\bibitem{rn19} J.-G. Wang, L.-L. Xu, and S.-J. Yang,
                      {\rm Ground-state phases of the spin-orbit-coupled spin-1 Bose gas in a toroidal trap},
                      Phys. Rev. A {\bf 96}, 033629 (2017).

\bibitem{rn20} N.-E. Guenther, P. Massignan, and A. L. Fetter,
                      {\rm Quantized superfluid vortex dynamics on cylindrical surfaces and planar annuli},
                      Phys. Rev. A {\bf 96}, 063608 (2017).

\bibitem{rn21} A. Roy and D. Angom,
                    {\rm Ramifications of topology and thermal fluctuations in quasi-2D condensates},
                    J. Phys. B {\bf 50}, 225301 (2017).

\bibitem{rn22} S. Eckel, A. Kumar, T. Jacobson, I.~B. Spielman, and G. K. Campbell,
                      {\rm A Rapidly Expanding Bose-Einstein Condensate: An Expanding Universe in the Lab},
                      Phys. Rev. X {\bf 8}, 021021 (2018).
  
\bibitem{hol1} S. K. Adhikari,
                     {\rm Dipolar Bose-Einstein condensate in a ring or in a shell},
                     Phys. Rev. A {\bf 85}, 053631 (2012).

\bibitem{hol2} K. Padavi\'c, K. Sun, C. Lannert, and S. Vishveshwara,
                     {\rm Physics of hollow Bose-Einstein condensates},
                     Europhys. Lett. {\bf 120}, 20004 (2017).

\bibitem{hol3} K. Sun, K. Padavi\'c, F. Yang, S. Vishveshwara, and C. Lannert,
                     {\rm Static and dynamic properties of shell-shaped condensates},
                     Phys. Rev. A {\bf 98}, 013609 (2018).

\bibitem{MCTDHB1} A. I. Streltsov, O. E. Alon, and L. S. Cederbaum,
                            {\rm Role of Excited States in the Splitting of
                            a Trapped Interacting Bose-Einstein Condensate by 
                            a Time-Dependent Barrier},  
                            Phys. Rev. Lett. {\bf 99}, 030402 (2007).

\bibitem{MCTDHB2} O. E. Alon, A. I. Streltsov, and L. S. Cederbaum,
                            {\rm Multiconfigurational time-dependent Hartree method for bosons: 
                            Many-body dynamics of bosonic systems},
                            Phys. Rev. A {\bf 77}, 033613 (2008).

\bibitem{MCTDH_BB} O. E. Alon, A. I. Streltsov, and L. S. Cederbaum,
                              {\rm Multiconfigurational time-dependent Hartree method for
                              mixtures consisting of two types of identical particles},
                              Phys. Rev. A. {\bf 76}, 062501 (2007).

\bibitem{BJJ} K. Sakmann, A. I. Streltsov, O. E. Alon, and L. S. Cederbaum, 
                    {\rm Exact Quantum Dynamics of a Bosonic Josephson Junction},
                    Phys. Rev. Lett. {\bf 103}, 220601 (2009).

\bibitem{MCTDHB_OCT} J. Grond, J. Schmiedmayer, and U. Hohenester,
                                  {\rm Optimizing number squeezing when splitting a mesoscopic condensate},
                                  Phys. Rev. A {\bf 79}, 021603(R) (2009).

\bibitem{Benchmarks} A. U. J. Lode, K. Sakmann, O. E. Alon, L. S. Cederbaum, and A. I. Streltsov, 
                                {\rm Numerically exact quantum dynamics of bosons with 
                                time-dependent interactions of harmonic type},
                                Phys. Rev. A {\bf 86}, 063606 (2012).

\bibitem{ML1} S. Kr\"onke, L. Cao, O. Vendrell, and P. Schmelcher,
                     {\rm Non-equilibrium quantum dynamics of ultra-cold atomic mixtures:
                     the multi-layer multi-configuration time-dependent Hartree method for bosons},
                     New J. Phys. {\bf 15}, 063018 (2013).

\bibitem{ML2} L. Cao, S. Kr\"onke, O. Vendrell, and P. Schmelcher,
                     {\rm The multi-layer multi-configuration time-dependent Hartree
                     method for bosons: Theory, implementation, and applications}, 
                     J. Chem. Phys. {\bf 139}, 134103 (2013).

\bibitem{MCTDHB_3D_stat} A. I. Streltsov, 
                                       {\rm Quantum systems of ultracold bosons
                                        with customized interparticle interactions},
                                        Phys. Rev. A {\bf 88}, 041602(R) (2013).

\bibitem{MCTDHB_3D_dyn} O. I. Streltsova, O. E. Alon, L. S. Cederbaum, and A. I. Streltsov,
                                        {\rm Generic regimes of quantum many-body dynamics of trapped 
                                        bosonic systems with strong repulsive interactions},
                                        Phys. Rev. A {\bf 89}, 061602(R) (2014).

\bibitem{Breaking} S. Klaiman, A. U. J. Lode, A. I. Streltsov, L. S. Cederbaum, and O. E. Alon, 
                           {\rm Breaking the resilience of a two-dimensional
                           Bose-Einstein condensate to fragmentation},                       
                           Phys. Rev. A {\bf 90}, 043620 (2014). 

\bibitem{Uwe} U. R. Fischer, A. U. J. Lode, and B. Chatterjee,
                     {\rm Condensate fragmentation as a sensitive measure of the quantum
                     many-body behavior of bosons with long-range interactions},
                     Phys. Rev. A {\bf 91}, 063621 (2015).

\bibitem{2D_Tun} R. Beinke, S. Klaiman, L. S. Cederbaum, A. I. Streltsov, and O. E. Alon,
                           {\rm Many-body tunneling dynamics of Bose-Einstein condensates and 
                           vortex states in two spatial dimensions},
                           Phys. Rev. A {\bf 92}, 043627 (2015).

\bibitem{Axel_MCTDHF_HIM} E. Fasshauer and A. U. J. Lode,
                                          {\rm Multiconfigurational time-dependent Hartree method for fermions: 
                                          Implementation, exactness, and few-fermion tunneling to open space},
                                          Phys. Rev. A {\bf 93}, 033635 (2016).

\bibitem{MCTDHB_spin} A. U. J. Lode,
                                   {\rm Multiconfigurational time-dependent Hartree method for
                                   bosons with internal degrees of freedom:     
                                   Theory and composite fragmentation
                                   of multicomponent Bose-Einstein condensates},
                                   Phys. Rev. A {\bf 93}, 063601 (2016).

\bibitem{Kaspar_n} K. Sakmann and M. Kasevich,
                            {\rm Single-shot simulations of dynamic quantum many-body systems},
                            Nat. Phys. {\bf 12}, 451 (2016).

\bibitem{ML3} L. Cao, V. Bolsinger, S. I. Mistakidis, G. M. Koutentakis,
                     S. Kr\"onke, J. M. Schurer, and P. Schmelcher,
                     {\rm A unified ab initio approach to the correlated quantum
                     dynamics of ultracold fermionic and bosonic mixtures},
                     J. Chem. Phys. {\bf 147}, 044106 (2017).

\bibitem{Higher_DIM} V. J. Bolsinger, S. Kr\"onke, and P. Schmelcher,
                                {\rm Beyond mean-field dynamics of ultra-cold bosonic atoms in higher dimensions:
                                facing the challenges with a multi-configurational approach},
                                J. Phys. B {\bf 50}, 034003 (2017).

\bibitem{Cami_NJP} C. L\'ev\^{e}que and L. B. Madsen,
                             {\rm Time-dependent restricted-active-space
                             self-consistent-field theory for bosonic many-body systems},
                             New J. Phys. {\bf 19}, 043007 (2017).

\bibitem{Axel_ar} S. E. Weiner, M. C. Tsatsos, L. S. Cederbaum, and A. U. J. Lode,
                          {\rm Phantom vortices: hidden angular momentum
                          in ultracold dilute Bose-Einstein condensates},
                          Sci Rep. {\bf 7}, 40122 (2017).

\bibitem{Axel_Cavity} A. U.  J. Lode and C. Bruder,
                               {\rm Fragmented Superradiance of a Bose-Einstein Condensate in an Optical Cavity},
                               Phys. Rev. Lett. {\bf 118}, 013603 (2017).

\bibitem{Dimensional_Cross} V. J. Bolsinger, S. Kr\"onke, and P. Schmelcher,
                                         {\rm Ultracold bosonic scattering dynamics off a repulsive barrier:
                                         Coherence loss at the dimensional crossover},
                                         Phys. Rev. A {\bf 96}, 013618 (2017).

\bibitem{2D_Dark} G. C. Katsimiga, S. I. Mistakidis, G. M. Koutentakis, P. G. Kevrekidis, and P. Schmelcher,
                            {\rm Many-body quantum dynamics in the decay of 
                            bent dark solitons of Bose-Einstein condensates},
                            New J. Phys. {\bf 19}, 123012 (2017).

\bibitem{Peter_Mix1} J.~M. Schurer, A. Negretti, and P. Schmelcher,
                               {\rm Unraveling the Structure of Ultracold Mesoscopic Collinear Molecular Ions},
                               Phys. Rev. Lett. {\bf 119}, 063001 (2017).

\bibitem{Peter_Mix2} J. Chen, J.~M. Schurer, and P. Schmelcher,
                              {\rm Entanglement Induced Interactions in Binary Mixtures},
                              Phys. Rev. Lett. {\bf 121}, 043401 (2018).

\bibitem{Cami_JPB} C. L\'ev\^{e}que and L. B. Madsen,
                            {\rm Multispecies time-dependent restricted-active-space self-consistent-field-theory
                            for ultracold atomic and molecular gases},
                            J. Phys. B {\bf 51}, 155302 (2018).

\bibitem{Axel_2018} R. Roy, A. Gammal, M. C. Tsatsos, B. Chatterjee, B. Chakrabarti, and A. U. J. Lode,
                            {\rm Phases, many-body entropy measures, and coherence of 
                            interacting bosons in optical lattices},
                            Phys. Rev. A {\bf 97}, 043625 (2018).

\bibitem{Alexej_2018} T. A. Elsayed and A. I. Streltsov,
                               {\rm Probing quantum states with momentum boosts},
                               Phys. Rev. A {\bf 98}, 013618 (2018).

\bibitem{PACK1} A. I. Streltsov and O. I. Streltsova, MCTDHB-Lab, version 1.5, 2015,
                        \href{http://www.mctdhb-lab.com}{http://www.mctdhb-lab.com}.

\bibitem{PACK2} A. I. Streltsov, L. S. Cederbaum, O. E. Alon, K. Sakmann,
                        A. U. J. Lode, J. Grond, O. I. Streltsova, S. Klaiman, and R. Beinke, 
                        The Multiconfigurational Time-Dependent Hartree for Bosons Package, 
                        version 3.x, \href{http://mctdhb.org}{http://mctdhb.org},
                        Heidelberg/Kassel (2006-present).

\bibitem{MCHB} A. I. Streltsov, O. E. Alon, and L. S. Cederbaum,
                       {\rm General variational many-body theory
                       with complete self-consistency for trapped bosonic systems},
                       Phys. Rev. A {\bf 73}, 063626 (2006).

\bibitem{cpl1990} H.-D. Meyer, U. Manthe, and L. S. Cederbaum,
                         {\rm The multi-configurational time-dependent Hartree approach},
                         Chem. Phys. Lett. {\bf 165}, 73 (1990).

\bibitem{jcp1992} U. Manthe, H.-D. Meyer, and L. S. Cederbaum,
                         {\rm Wave-packet dynamics within the multiconfiguration Hartree framework:
                         General aspects and application to NOCl},
                         J. Chem. Phys. {\bf 97}, 3199 (1992). 

\bibitem{review_Dieter} M. H. Beck, A. J\"ackle, G. A. Worth, and H. D. Meyer,
                                 {\rm The multiconfiguration time-dependent Hartree (MCTDH) method: 
                                 a highly efficient algorithm for propagating wavepackets},
                                Phys. Rep. {\bf 324}, 1 (2000).

\bibitem{book_MCTDH} {\it Multidimensional Quantum Dynamics: MCTDH Theory and Applications}, 
                                 edited by H.-D. Meyer, F. Gatti, and G. A. Worth
                                 (Wiley-VCH, Weinheim, 2009).

\bibitem{ML-MCTDH} H. Wang and M. Thoss,
                               {\rm Multilayer formulation of the multiconfiguration time-dependent Hartree theory},
                               J. Chem. Phys. {\bf 119}, 1289 (2003).

\bibitem{ML-Manthe} U. Manthe,
                               {\rm A multilayer multiconfigurational time-dependent Hartree approach for quantum
                                dynamics on general potential energy surfaces},
                                J. Chem. Phys. {\bf 128}, 164116 (2008).

\bibitem{ML-Oriol} O. Vendrell and H.-D. Meyer,
                           {\rm Multilayer multiconfiguration time-dependent Hartree method:
                           Implementation and applications to a Henon-Heiles Hamiltonian and to pyrazine},
                           J. Chem. Phys. {\bf 134}, 044135 (2011).

\end{thebibliography}
\end{document}